\documentclass[%
 aip,
 amsmath,amssymb,
 reprint,%
]{revtex4-1}

\usepackage{graphicx}
\usepackage{dcolumn}
\usepackage{bm}

\usepackage[utf8]{inputenc}
\usepackage[T1]{fontenc}
\usepackage{mathptmx}
\usepackage{etoolbox}

\usepackage{amsmath}
\usepackage{amsfonts}
\usepackage{amssymb}
\usepackage{amsbsy}
\usepackage{wrapfig}
\usepackage{times}
\usepackage{multirow}
\usepackage{footmisc}
\usepackage{rotating}
\usepackage{subfigure}
\usepackage{color}
\usepackage{setspace}
\usepackage{hyperref}
\usepackage{natbib}
\bibpunct{(}{)}{;}{a}{,}{,}

\newlength{\bigfigheight}
\newlength{\smallfigheight}

\makeatletter
\def\@email#1#2{%
 \endgroup
 \patchcmd{\titleblock@produce}
  {\frontmatter@RRAPformat}
  {\frontmatter@RRAPformat{\produce@RRAP{*#1\href{mailto:#2}{#2}}}\frontmatter@RRAPformat}
  {}{}
}%
\makeatother
\begin{document}

\preprint{AIP/123-QED}

\title[]{Direct statistical simulation of the Lorenz63 system}
\author{Kuan Li}
\email{kuan.li.81@gmail.com.}
 \affiliation{Department of Applied Mathematics, University of Leeds, Leeds, LS2 9JT, UK.}
\author{J.B. Marston}%
\affiliation{Department of Physics, Box 1843, Brown University, Providence, Rhode Island 02912-1843, USA, \\
Brown Theoretical Physics Center, Brown University, Providence, Rhode Island 02912-S USA.
}%
\author{Saloni Saxena}%
\affiliation{Department of Physics, Box 1843, Brown University, Providence, Rhode Island 02912-1843, USA.}%
\author{Steven M. Tobias}
\affiliation{Department of Applied Mathematics, University of Leeds, Leeds, LS2 9JT, UK.}%

\date{\today}

\maketitle

%

\section{Abstract}
We use direct statistical simulation (DSS) to find the low-order statistics of the well-known dynamical system, the Lorenz63 model. Instead of accumulating statistics from numerical simulation of the dynamical system, we solve the equations of motion for the statistics themselves after closing them by making several different choices for the truncation.  Fixed points of the statistics are obtained either by time evolving, or by iterative methods.  Statistics so obtained are compared to those found by the traditional approach.  

\section{Introduction}

Chaotic dynamical systems can be characterized in a number of different ways such as Lyapunov exponents, rare events, and low-order statistics.  Appropriate methods should be tailored to the particular type of information that is being sought.  In this paper we explore the low-order equal-time statistics of the celebrated dynamical system, the Lorenz63 model. To find the low-order cumulants we employ the method of direct statistical simulation (DSS) that works directly with the equations of motion for the statistics themselves, a program that Lorenz himself described \citep{lorenz67} in the context of numerical weather prediction.  Because nonlinearities couple successive orders of the hierarchy of cumulants together, it is necessary to make closure approximations that truncate the equations of motion.  We examine the reliability of several closure approximations by comparison to statistics obtained by the traditional method of time integrating the dynamical system. 

The chaotic Lorenz63 model is ideal for a study of the capabilities and limitations of DSS because it is highly non-linear -- more nonlinear in fact than many anisotropic and inhomogeneous problems in fluid dynamics, such as jets that can be viewed as perturbations to a mean-flow (see for instance \citep{mqt2019}). The strong quadratic nonlinearity tests the limits of different closure approximations.  In addition, the low dimensionality of the Lorenz63 system makes it possible to query all of the low-order equal time statistics. 

Our results extend previous work \citep{allawala_2016} in several ways. First we use symbolic manipulation with {\it SymPy} to generate a dense set of equations of motion for the statistics for the dynamical system, instead of relying on general arrays to encode coefficients. Second we consider the CE2.5 level of approximation that is closely related to the eddy-damped quasi-normal Markovian (EDQNM) approximation \citep{mqt2019}. Third and perhaps most importantly, we attempt to use iterative methods to find the fixed points of the truncated cumulant equations of motion that represent the statistical steady state. 

The organization of the rest of the paper is as follows. In Section \ref{cumulants} we introduce the equal-time cumulants, closure approximations, and methods for finding fixed points. Section \ref{lz63} discusses our results for the Lorenz63 attractor. Some conclusions are presented in Section \ref{conclusion}.  

\section{Cumulant statistics of low-order dynamical systems}
\label{cumulants}

In this section we briefly summarize how the low-order statistics of the low-order dynamical systems can be approximated using direct statistical simulation.

Low-order dynamical systems are reduced models often used to represent the characteristic behaviours of the fluid dynamical related problems. These models are derived either via a Galerkin truncation of the partial differential equation (PDE) system, e.g., see \citep{holmes_etal_2012}, \citep{Maasch_1990} or via a normal form analysis \citep[see e.g.][]{Tobias_1995}. 

We consider a nonlinear system, whose governing equation are generically represented by a set of ordinary differential equations with up to quadratic nonlinearities;  the $i$-th component reads
\begin{equation}
	d_t x_i =  \sum_j {\cal L}_{ij} \ x_j +   \sum_{j,k} {\cal Q}_{ijk} \ x_j x_k + f_i,
\label{govEq}
\end{equation}
where the coefficient of the quadratic nonlinear interaction is given by ${\cal Q}_{ijk}$ and ${\cal L}_{ij}$ is for the linear term.  Here we also include the possibility of a stochastic forcing, $f_i$,  that is assumed to be an independent Gaussian, $f_i \sim{\cal N}(\mu_i,  \sigma_i^2)$, is introduced to synthesize the unmodelled physical processes, where $\mu_i$ and $\sigma_i^2$ are the statistical mean and variance of $f_i$, respectively \citep[see e.g.][]{allawala_2016}.

Direct statistical simulation describes the evolution of dynamical systems in probability space. In this framework, the unknown field, $x_i$, is treated as a random variable with the associated probability density function (PDFs) represented by a series of statistics, namely cumulants \citep{Kendall_1987}. The random variable, $x_i$, which consists of the coherent component, $C_{x_i}$, and the non-coherent counterpart, $\delta x_i$, i.e., 
	\begin{equation}
			x_i = C_{x_i} + \delta {x_i},
		\label{rede}
	\end{equation}
is required to satisfy the Reynolds averaging rules, e.g.,
\begin{eqnarray}
&&C_{x_i} = \langle x_i \rangle, \ \ \ \ \ 0 = \langle \delta x_i \rangle = 0,   \nonumber \\
&&\langle x_i C_{x_j} \rangle = \langle (C_{x_i}+\delta x) C_{x_j} \rangle = \langle  C_{x_i} C_{x_j} \rangle
\end{eqnarray}
where the statistical average, noted as $\langle \ \rangle$, is assumed to be the ensemble average in this study. In the cumulant hierarchy, the first three comulants, $C_{x_i}$, $C_{x_ix_j}$ and $C_{x_ix_jx_k}$, are identical to the statistical central moments. The fourth and higher cumulants, unlike the second and third terms, are arbitrarily zeros for a Gaussian distribution, whilst the fourth centred moment does not. In this paper, we explicitly use the fourth cumulant, which is defined as
\begin{eqnarray}
C_{x_ix_jx_kx_l} &=& \langle \delta x_i \delta x_j \delta x_k \delta x_l \rangle \nonumber \\
&-& C_{x_ix_j} C_{x_kx_l}- C_{x_ix_k} C_{x_jx_l} - C_{x_jx_k} C_{x_ix_l}. 
\end{eqnarray}

The governing equations of the cumulants can be derived directly from the dynamical equation (\ref{govEq}) or via the Hopf functional approach \citep{frisch_1995}. In this study, we choose the first approach and use the existing software in {\it Python} to derive the cumulant equations. This package has been developed by the authors for studying the low-order dynamo problems \citep{Li_etal_2021} and available online (\url{https: //github.com/Kuan-Li-Math-Geo/dss_low-order.git}). The first three cumulant equations are introduced in Eqs. (2.6--2.8) in \citet{Li_etal_2021}, where for the quadratically nonlinear system (\ref{govEq}) considered here the terms with the cubic coefficients do not appear.

The expansion of the nonlinear dynamical equation (\ref{govEq}) leads to an infinite hierarchy of coupled equations, i.e., for the quadratic system, the $i$th cumulant equation always involves the $i+1$th cumulant. A proper statistical closure must be chosen to truncate the cumulant expansion at the lowest possible order.  We truncate the cumulant equations using three different truncation rules, namely CE2, CE2.5 and CE3. The CE2 is ideal for studying the dynamical systems with the state vector satisfying or close to Gaussian distributions, where the cumulant hierarchy is truncated at second order and all higher order terms greater than two are neglected \citep[see e.g.][]{mqt2019}. The CE3 rule is often applied to truncate the dynamical systems with strong asymmetry (skewness) or long tails (flatness) in PDFs as we will see in \$\ref{lz63}. In CE3 approximations, the set of cumulant equations is truncated at the third order, whilst the fourth order cumulant is set to zero $C_{x_ix_jx_kx_l}=0$, \citep{orszag_1970}. The effects of the fourth order cumulants are further modelled by a diffusion process, $-C_{x_ix_mx_n}/\tau_d$, with the parameter, $\tau_d>0$, known as the eddy damping parameter \citep{mqt2019}, e.g., refer to Eqs. (2.6, 2.7 and 2.10) in \citet{Li_etal_2021}. The damping parameter, $\tau_d$, measures the time scale of the noncoherent components of the flow resulting from their nonlinear interaction. In the highly chaotic regimes of the dynamical system, $\tau_d$ is expected to be a small number. The third cumulant, $C_{x_ix_jx_k}$, evolves much more rapidly in time as compared with the first and second cumulant. A further simplification of the third order cumulant equations can be made, which leads to the so-called CE2.5 approximation \citep{mqt2019} and \citep{allawala_2020}. In CE2.5, the third cumulant are determined by solving the diagnostic components of the third order equation with $d_t C_{x_ix_mx_n}=0$ and with all first order cumulant setting to zero, $C_{x_i}=0$, to further speed up the computation, e.g., see Eqs. (2.6, 2.7 and 2.11) in \citet{Li_etal_2021}.




The low-order statistics governed by the cumulant equations evolves much smoother in space and time than the instantaneous field, $x_i$. As forward evolving the cumulant equations,  the solution settles in statistical equilibrium states, which is invariant in time.  This approach will determine stable solutions to the cumulant equations. However, many of the time-invariant solutions of the cumulant system are either unstable as we integrate CE2/2.5/3 equations in time or statistically non-realizable. Realizability is satisfied only if the second cumulant satisfies the Cauchy–Schwarz inequality, $C_{x_i}^2C_{x_j}^2 \ge C_{x_ix_j}^2$. We use three different methods to study the stability and statistical realisability of the fixed points by 1) forward evolving the cumulant equations, 2) directly solving the spatial components of the (algebraic) cumulant equations, where the temporal components are set to zero and 3) solving an inverse problem \citep{Nocedal_06, Karft_1988}. The definition of the misfit functional, ${\cal J}$ is found in Eq. (2.12) in \citep{Li_etal_2021}.

\section{Lorenz63 system}
\label{lz63}

The governing equation of the Lorenz63 system \citep{Lz_63} is obtained from the truncation of the Galerkin discretization of the atmospheric convection model at the lowest order, where the evolution equations are given by
\begin{eqnarray}
	    \left(d_t + P_r     \right) x &=& P_r y  + f_x \nonumber \\
        \left(d_t + 1        \right) y &=& R_a x - x z + f_y \nonumber\\
        \left(d_t + \beta \right) z &=& x y + f_z.
\label{Lz63eq}
\end{eqnarray}
The control parameters are the Prandtl number, $P_r$, the (relative) Rayleigh number, $R_a$ and the geometric factor, $\beta$. The unknown functions, $x, y$ and $z$ represent the velocity, horizontal temperature variation and  vertical temperature variation, respectively. The stochastic force, $f_{x,y,z}$, is introduced to synthesize the unmodelled physical processes \citep{allawala_2016}, where $f_{x,y,z}$ is assumed to be independent Gaussian variable, $f_{x,y,z}\sim{\cal N}(\mu_{x,y,z}, \sigma^2_{x,y,z})$ with the mean and variance given by $\mu_{x,y,z}$ and $\sigma^2_{x,y,z}$.

We use the {\it Python} package to derive the cumulant equations in (\ref{Lz63eq}) and obtain the low-order cumulant approximations of Lorenz63 system, where the first and the second order equations read
\begin{eqnarray}
\left(d_t+P_r \right) C_x &=&   P_r C_y   + \mu_x, \nonumber \\
\left(d_t+ 1   \right) C_y &=&   -C_x C_z - C_{xz} + R_a C_x   + \mu_y, \nonumber \\
\left(d_t+\beta \right)C_z &=&   C_x C_y + C_{xy} + \mu_z,
\label{Lz63_ce1}
\end{eqnarray}
and
\begin{eqnarray}
\left(d_t+ 2 P_r \right) C_{xx} &=&   2P_r C_{xy} + 2 \sigma_x^2, \nonumber\\
\left(d_t+2 \right) C_{yy} &=&   -2 C_x C_{yz} - 2 C_{xy} C_z + 2 R_a C_{xy} \nonumber \\ 
&-& 2 C_{xyz}  + 2\sigma_y^2, \nonumber\\
\left(d_t+2 \beta \right) C_{zz} &=&   2 C_xC_{yz} + 2 C_{xyz} + 2 C_{xz} C_y + 2\sigma_z^2 \nonumber\\
\left(d_t+P_r + 1 \right) C_{xy} &=&   -C_x C_{xz} - C_{xx} C_z + R_a C_{xx} - C_{xxz} \nonumber \\ &+& P_r C_{yy},  \nonumber\\ 
\left(d_t+P_r + \beta  \right) C_{xz} &=&   C_x C_{xy} + C_{xx} C_y + C_{xxy} + P_r C_{yz},   \nonumber\\
\left(d_t+\beta + 1 \right) C_{yz} &=&   C_x C_{yy} - C_x C_{zz} + C_{xy} C_y + C_{xyy} \nonumber \\ &-& C_{xz} C_z + R_a C_{xz} - C_{xzz}.
\label{Lz63_ce2}
\end{eqnarray}
The third order equations have a complicated form and are detailed in Eqs. (\ref{Lz63_ce3}) in Appendix (\ref{Lz63_CE3}). The cumulant equations are further truncated according to CE2/2.5/3 truncation rules, respectively.

In this study, we vary the Rayleigh number, $R_a$, and the stochastic force, $f_{x,y,z}$, to control the dynamics of Lorenz63 system, whilst the Prandtl number, $P_r=10$, and the geometric factor, $\beta=8/3$, remain the same as those chosen by \citet{Lz_63} for all cases.

We focus our study on the chaotic regime of Lorenz63. We find two physical mechanisms that are able to drive the dynamics of Lorenz63 to the chaotic states. If the Rayleigh number is supercritical, i.e., $R_a$ is greater than a critical value, $R_{a_c}$, the Lorenz63 system spontaneously evolves towards the chaotic state. The chaotic solutions of Lorenz63 can also be obtained in the subcritical branch for $R_a<R_{a_c}$ by employing a strong external stochastic force, $f_{x,y,z}$. We compare the low-order statistics of Lorenz63 system obtained by DNS and DSS in the chaotic regime and show the effectiveness of the low-order cumulant approximation for describing the chaos of the Lorenz63 system.

\subsection{Lorenz63 system in the chaotic regime}

In the absence of external forcing, the Lorenz63 system (\ref{Lz63eq}) always evolves towards a steady state for $1<R_a<R_{a_c}$. Depending on the choice of the initial condition, the steady solution of Lorenz63 converges to one of two stable attractors at
\begin{equation}
{\bf X}_{{\cal F_{\pm}}} = (\pm\sqrt{\beta(R_a-1)}, \ \ \pm\sqrt{\beta(R_a-1)}, \ \ R_a-1).
\label{fixPLz}
\end{equation}
The critical Rayleigh number is determined by the Prandtl number, $P_r$ and the geometric factor, $\beta$, where $R_{a_c}$ is found to be approximately $25$ for $P_r=10$ and $\beta=8/3$. In the phase space, two stable attractors are non-connected but accessible with equal probability, due to the reflective symmetry, $x\rightarrow -x$ and $y\rightarrow -y$. 

\begin{figure*}[htp]
\centering
\subfigure[]
	{
		\includegraphics[width=0.15\hsize]{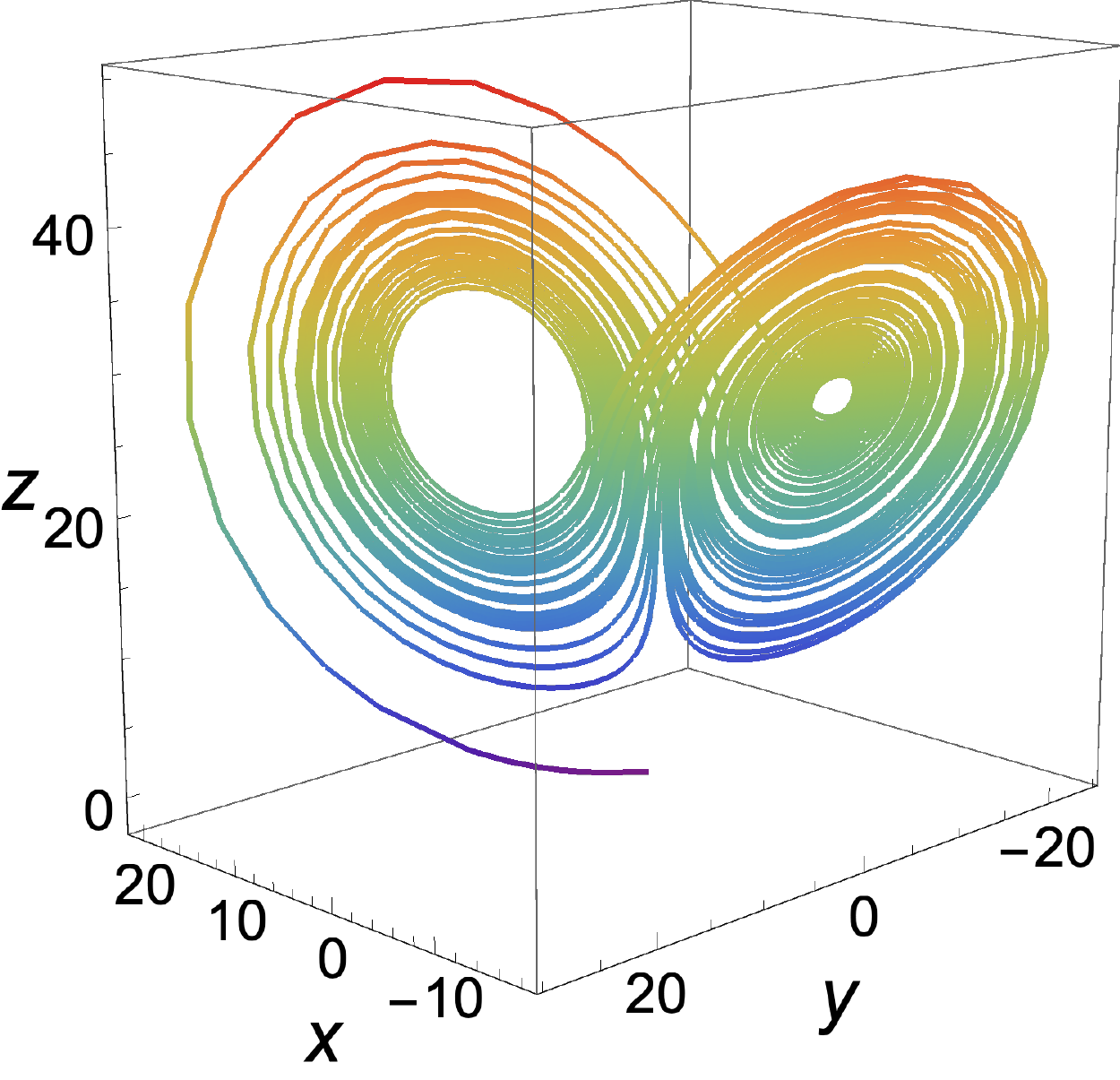}
	}
\subfigure[]
	{
		\includegraphics[width=0.18\hsize]{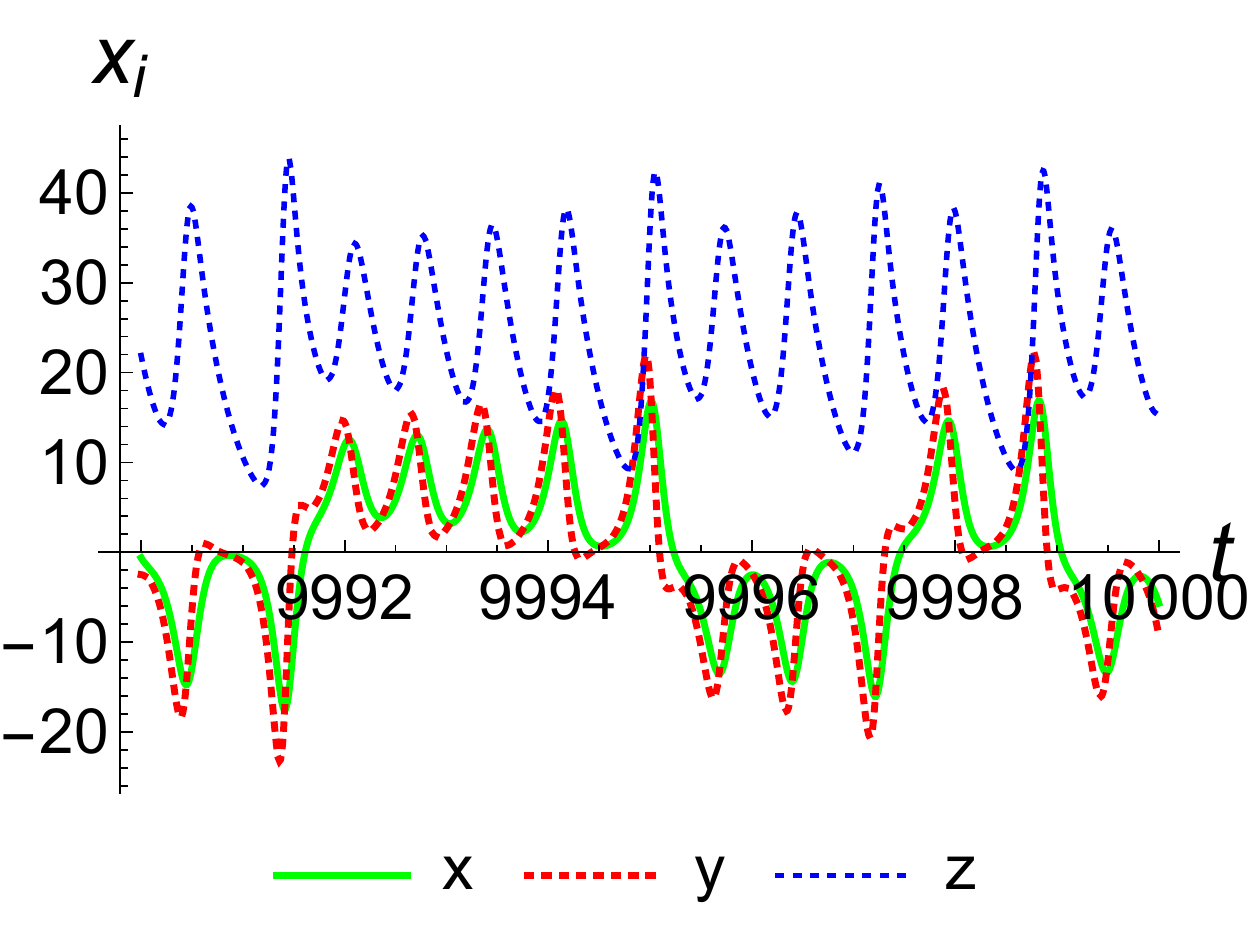}
	}
\subfigure[]
	{
		\includegraphics[width=0.18\hsize]{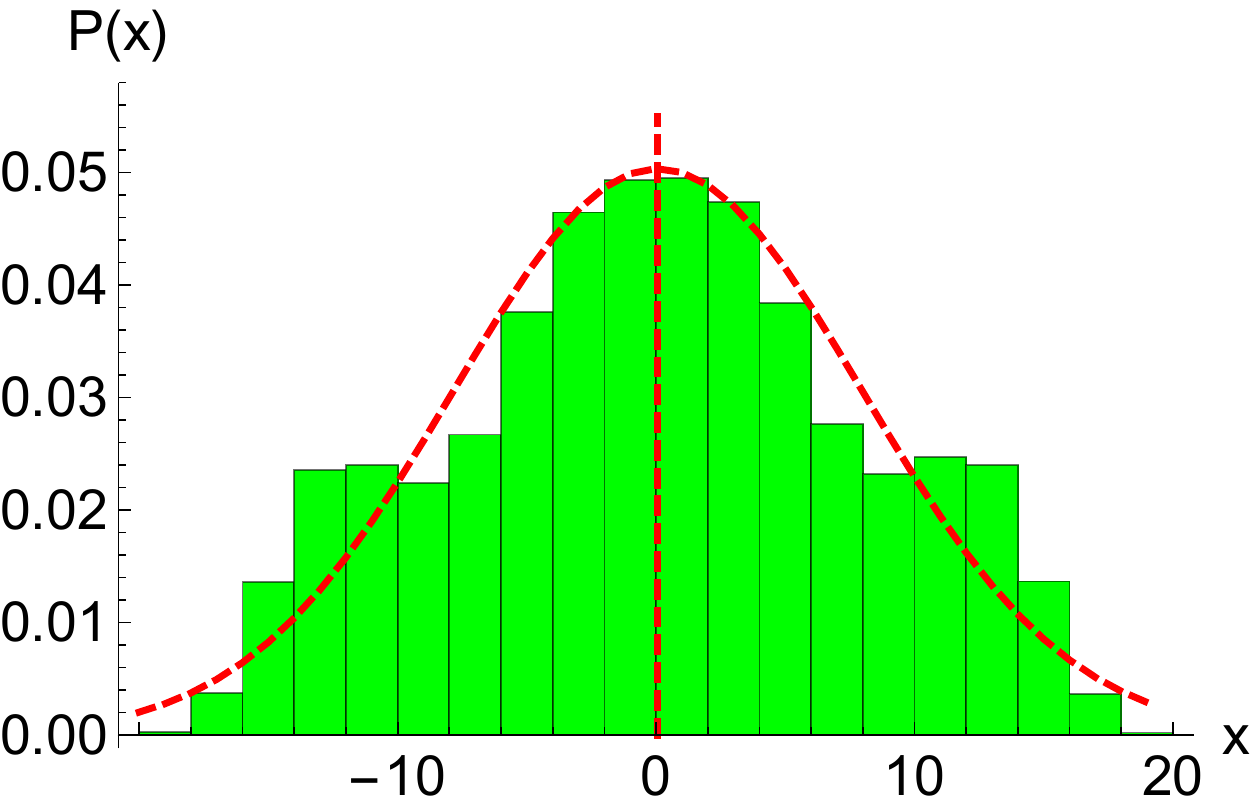}
	}
\subfigure[]
	{
		\includegraphics[width=0.18\hsize]{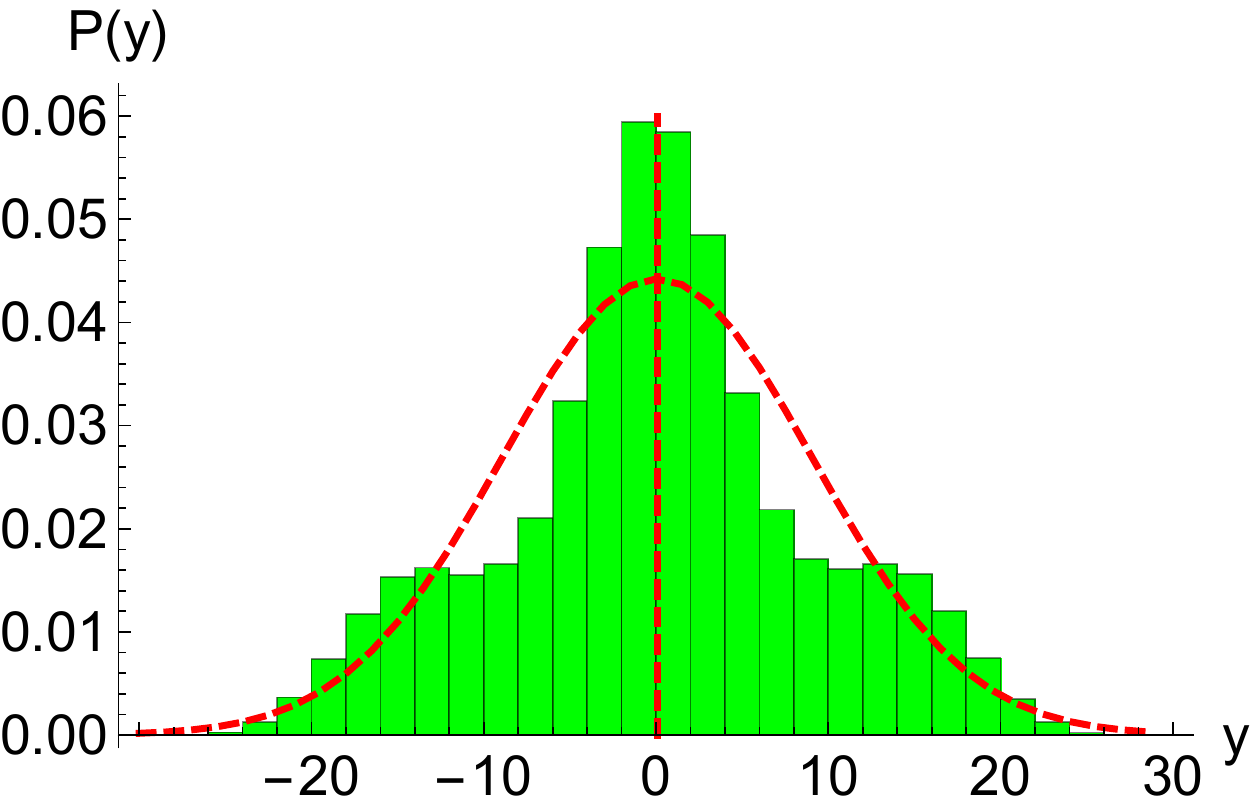}
	}
\subfigure[] 
	{
		\includegraphics[width=0.18\hsize]{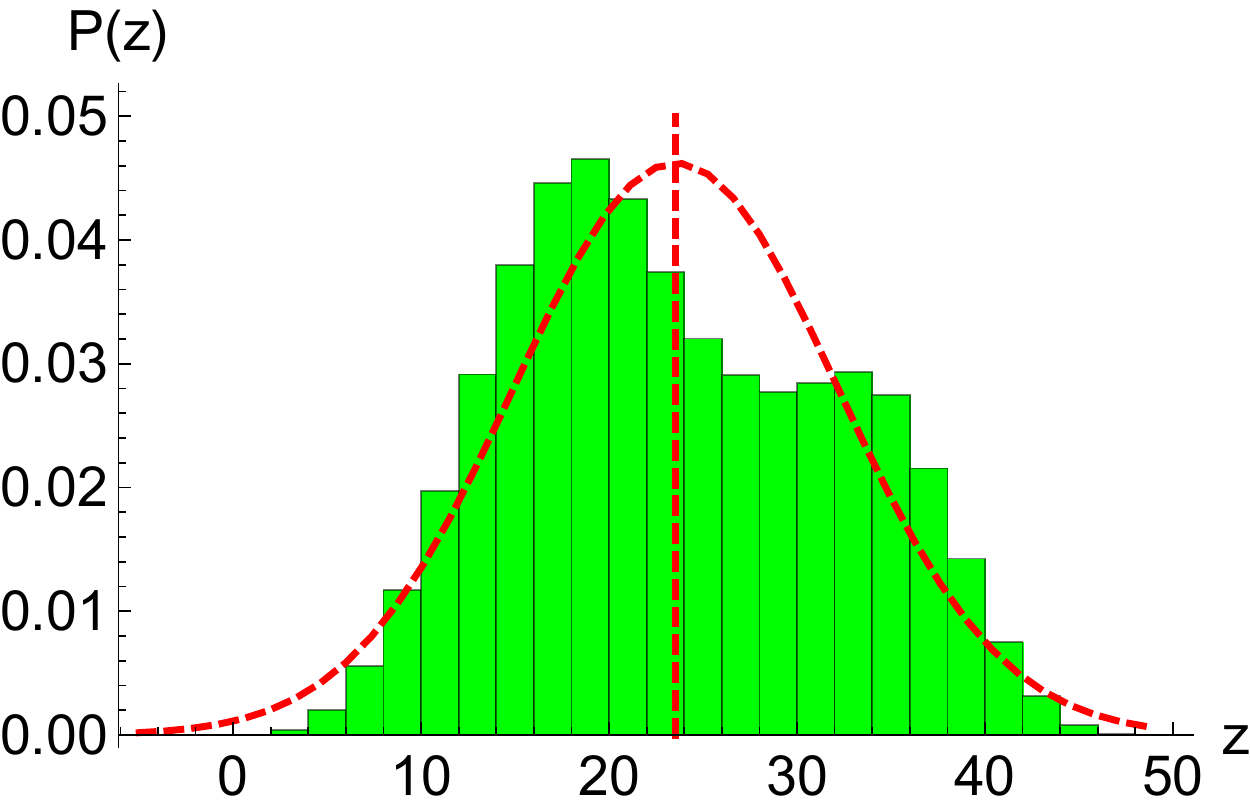}
	}
\caption{The illustration of the trajectory, time series and the PDFs of Lorenz63 in the chaotic state for $R_a=28$ and $f_{x,y,z}=0$, where the green histograms stand for the PDFs of $x, y$ and $z$ and the dashed red curves are for the Gaussian distribution with the same mean and variance as for DNS in (c)--(e).}
\label{Lz63_pt2}
\end{figure*}

In the supercritical regime for $R_a>R_{a_c}$, the attractors, ${\bf X}_{{\cal F_{\pm}}}$, become unstable and the trajectory of Lorenz63 repelled by the unstable (`strange') attractors oscillates irregularly in the phase space. Shown in Fig. (\ref{Lz63_pt2})
is the typical solution of the Lorenz63 system in the chaotic state for $R_a=28$. For this parameter set, the characteristic time scales of the Lorenz63 system are found to be $t_w \approx 3.5$ and $t_c \approx 0.75$, where $t_w$ is the average time for the trajectory to stay on one wing of the Lorenz butterfly and $t_c$ is the average time for the trajectory to orbit one of the wings. The probability distributions, ${\cal P} (x)$, ${\cal P} (y)$ and ${\cal P} (z)$, that are shown in Fig. (\ref{Lz63_pt2}c--e), are non-Gaussian, which indicate the importance of the higher order statistics in the approximation of ${\cal P} (x)$, ${\cal P} (y)$ and ${\cal P} (z)$, where the green histograms stand for probability distributions, ${\cal P} (x)$, ${\cal P} (y)$ and ${\cal P} (z)$, and the dashed red curves are for the Gaussian distribution with the same mean and variance for comparison purposes. The locations of the strange attractors in $x$ and $y$ coordinates appear as the small peaks on the left and right hand side of ${\cal P}(x)$ and ${\cal P}(y)$; the central peak that has the largest amplitude results from the rapid oscillation of the trajectory between two strange attractors. The PDF of $z$ is bimodal.

\begin{figure*}[htp]
\centering
\subfigure[]
	{
		\includegraphics[width=0.18\hsize]{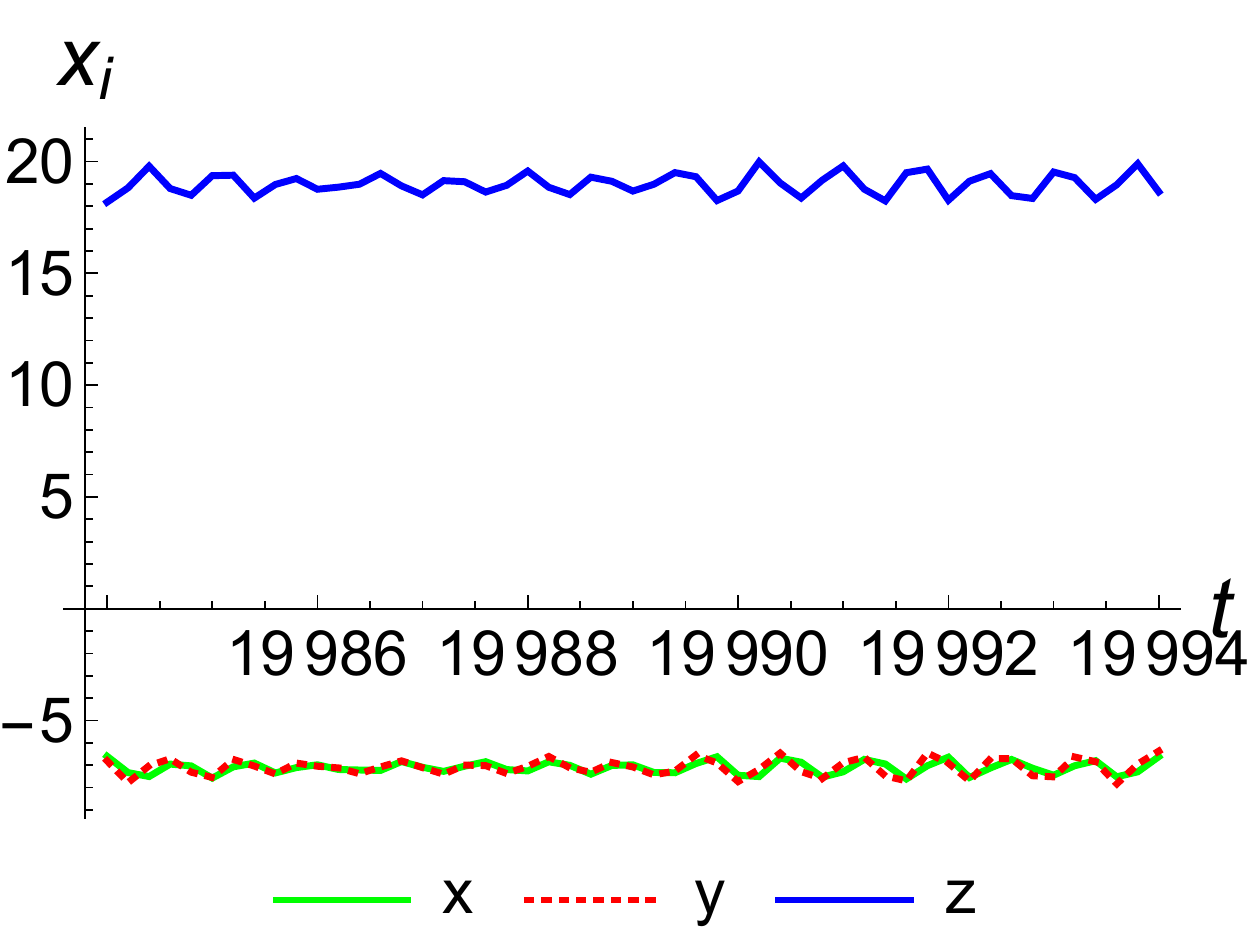}
	}
\subfigure[]
	{
		\includegraphics[width=0.18\hsize]{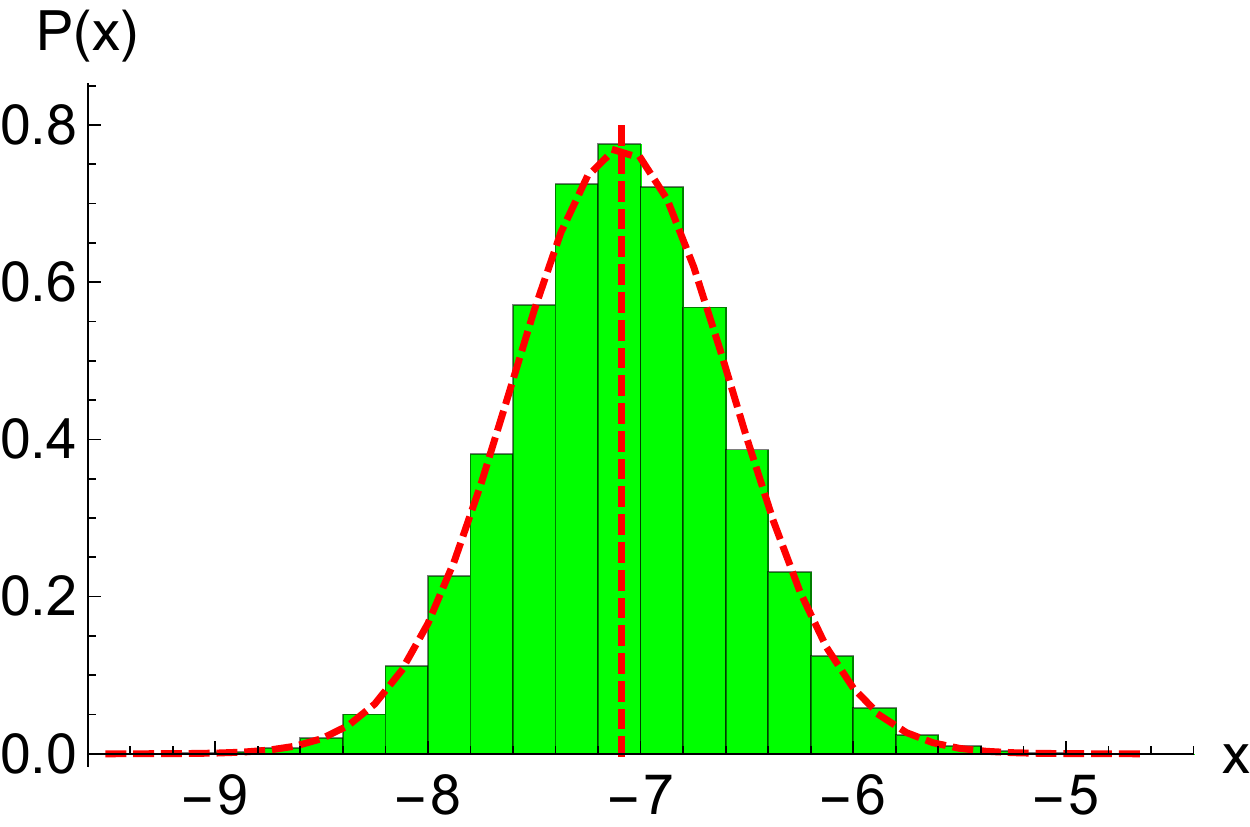}
	}
\subfigure[]
	{
		\includegraphics[width=0.18\hsize]{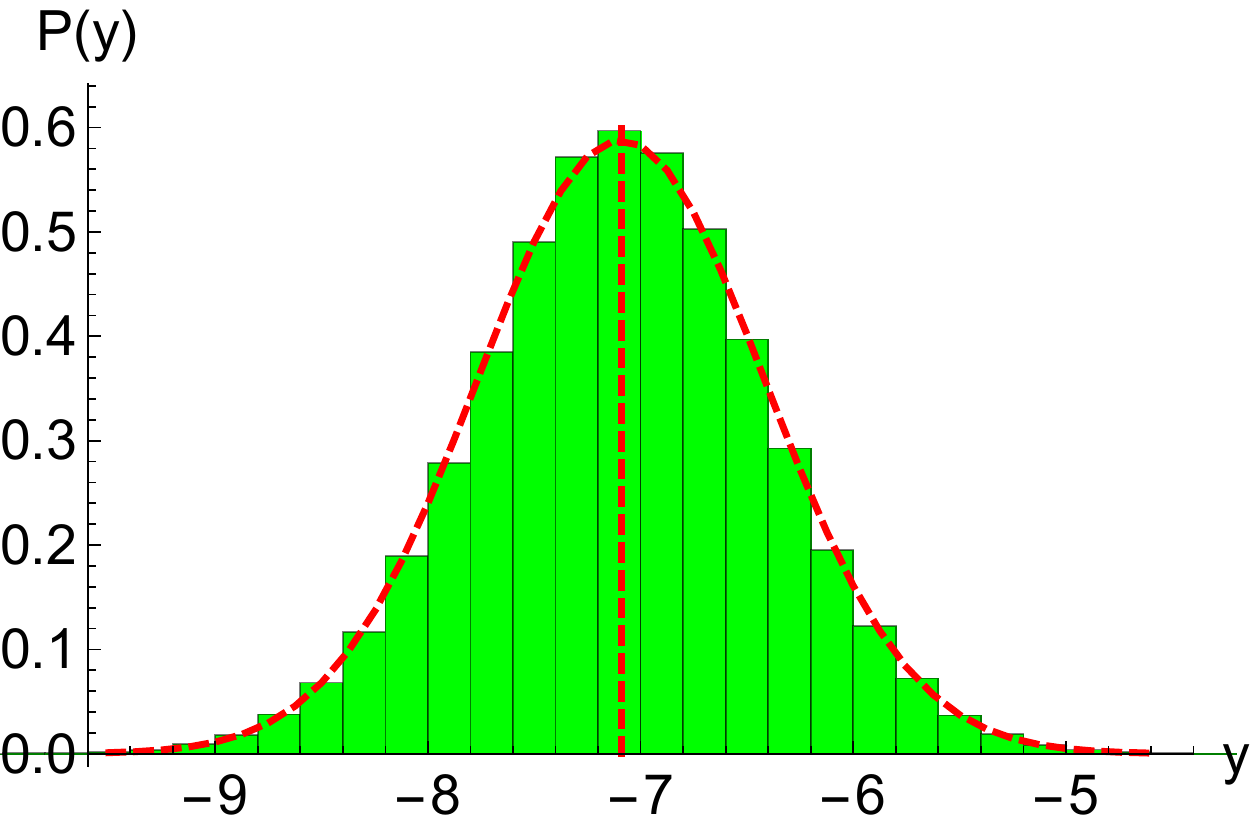}
	} 
\subfigure[]
	{
		\includegraphics[width=0.18\hsize]{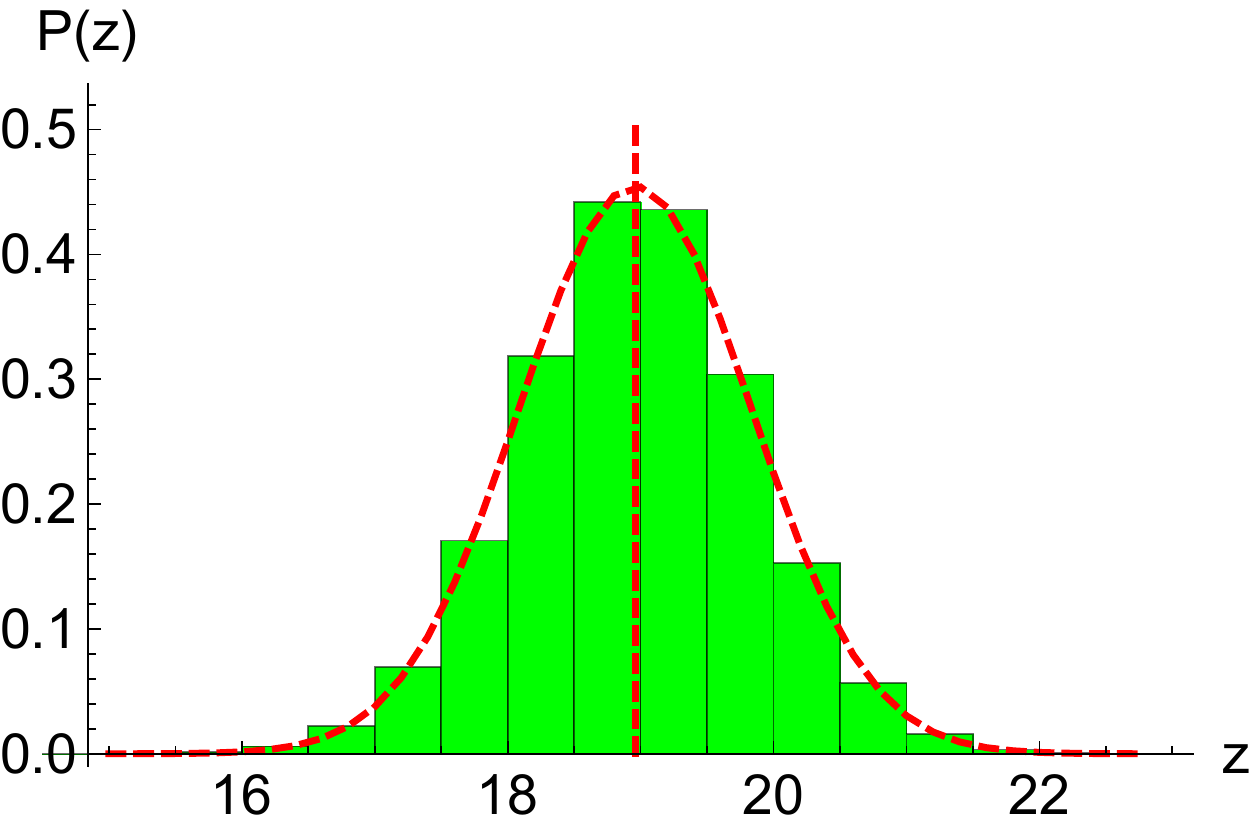}
	}
\caption{The illustration of the solution of Lorenz63 that oscillates about ${\bf X}_{{\cal F}_-} = [-7.1, -7.1, 19]$ for $R_a=20$ and $f_{x,y,z} \sim {\cal N}(0, \ 0.1)$, where the green histograms stand for the PDFs of $x, y$ and $z$ and the dashed red curves are for the Gaussian distribution with the same mean and variance as for DNS (b)--(d).}
\label{Lz_steady}
\end{figure*}

\begin{figure*}[htp]
\centering
\subfigure[]
	{
		\includegraphics[width=0.18\hsize]{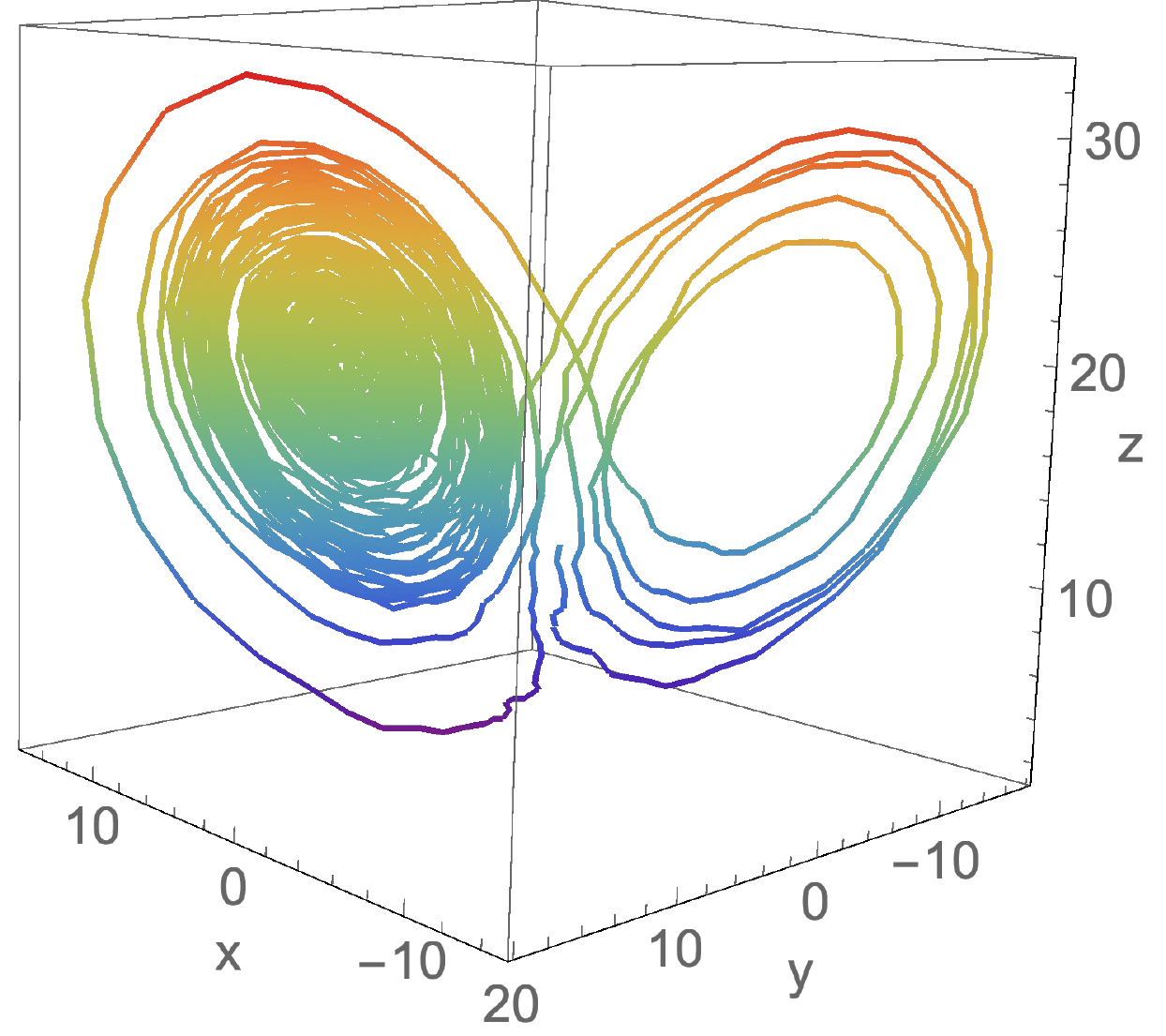}
	}
\subfigure[]
	{
		\includegraphics[width=0.18\hsize]{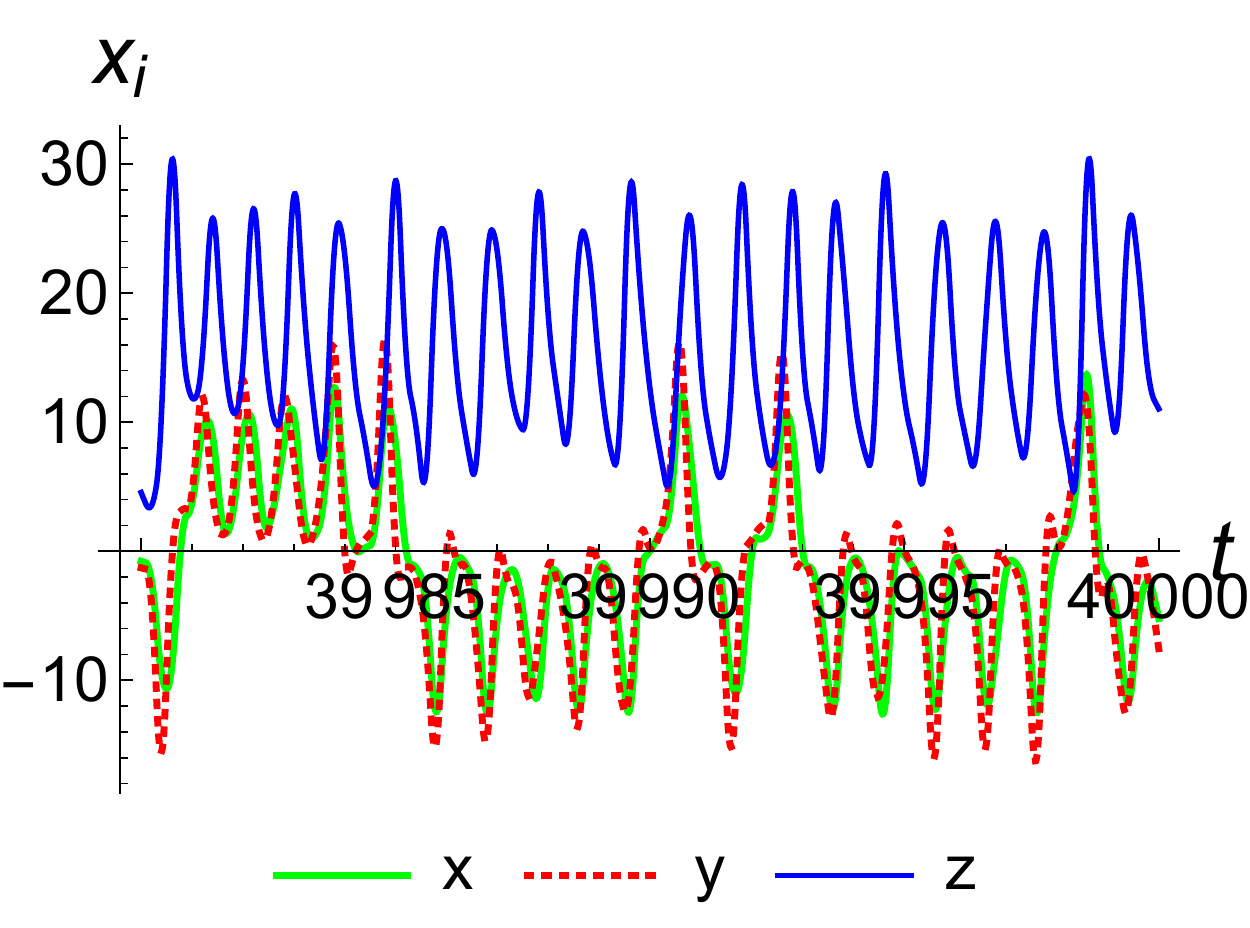}
	}
\subfigure[]
	{
		\includegraphics[width=0.18\hsize]{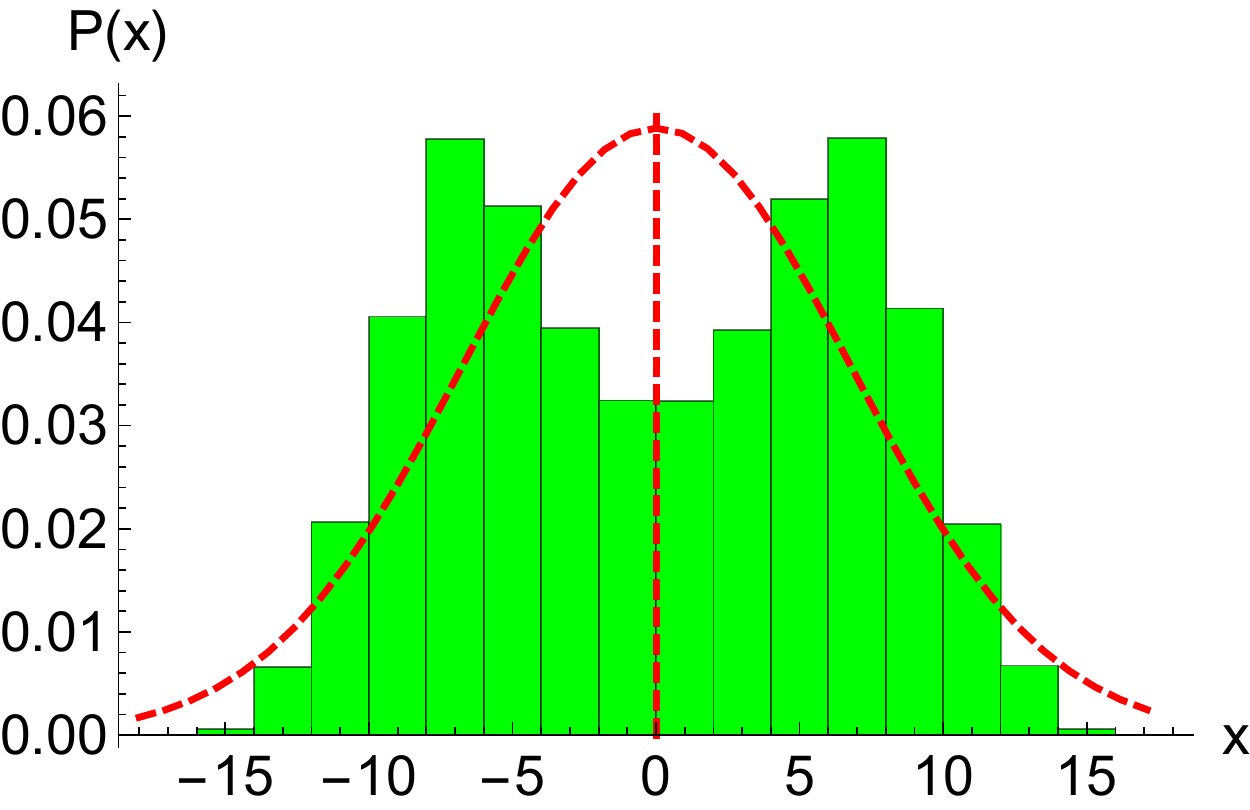}
	}
\subfigure[]
	{
		\includegraphics[width=0.18\hsize]{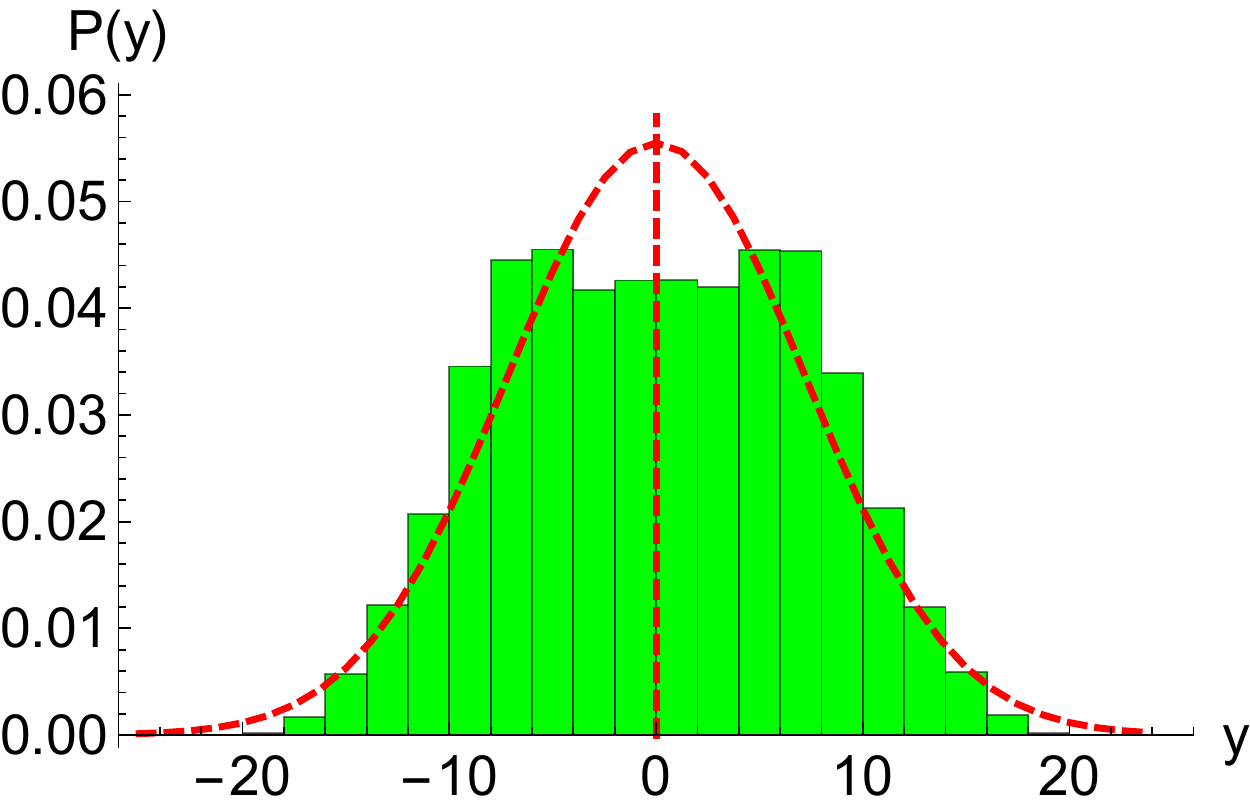}
	} 
\subfigure[]
	{
		\includegraphics[width=0.18\hsize]{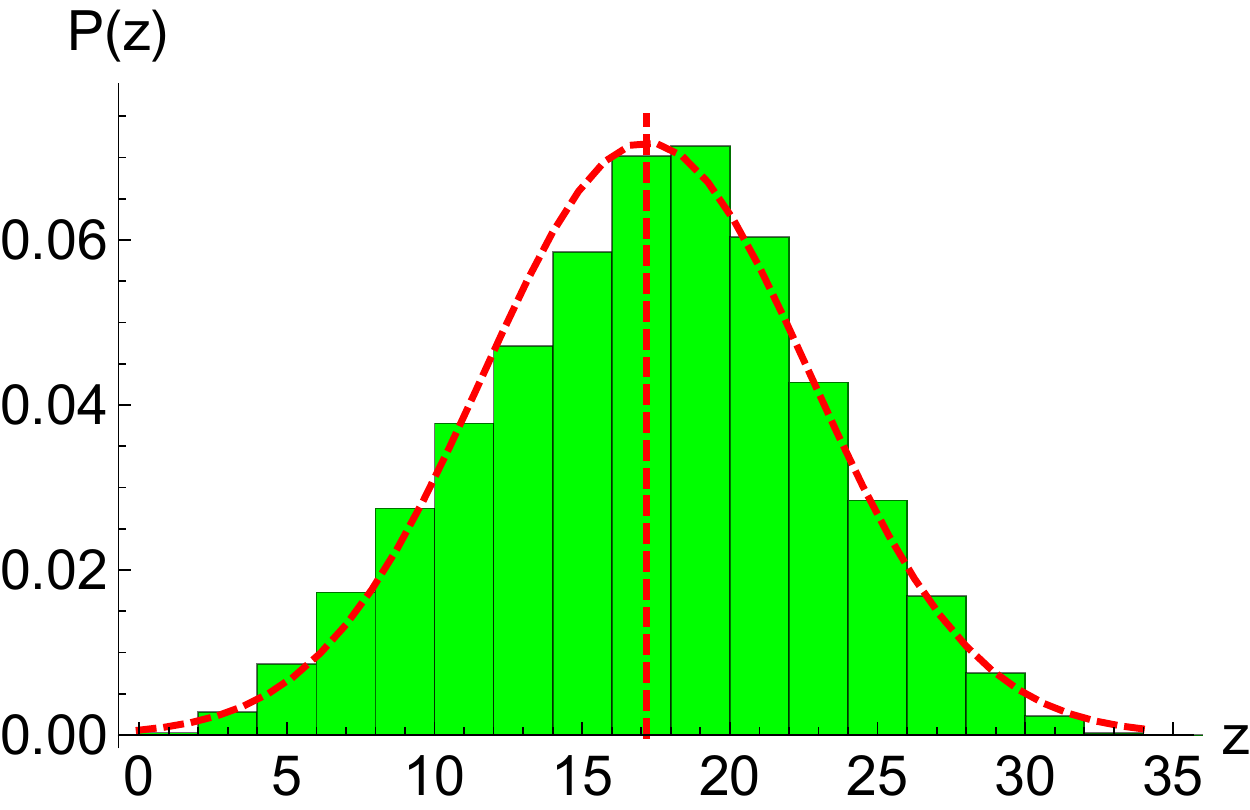}
	}
\caption{The illustration of the trajectory, time series and the PDFs of Lorenz63 in the chaotic state for $R_a=20$ and $f_{x,y,z}\sim{\cal N}(0, \ 2)$, where the green histograms stand for the PDFs of $x, y$ and $z$ and the dashed red curves are for the Gaussian distribution with the same mean and variance as for DNS in (c)--(e).}
\label{Lz63_pt3}
\end{figure*}

In the subritical regime for $1<R_a < R_{a_c}$, the stable attractor of Lorenz63 may become unstable as we increase the external stochastic force, $f_{x,y,z}$. For small random forces, i.e., $\sigma^2_{x,y,z}\ll {\cal O}(1)$, the trajectory of Lorenz63 oscillates randomly around one of two steady solutions at ${\bf X}_{\cal F_{\pm}}$ with the PDFs, $P(x)$, $P(y)$ and $P(z)$ in Gaussian, where ${\bf X}_{\cal F_{\pm}}= [\pm7.1, \pm 7.1, 19]$, e.g., see Fig. (\ref{Lz_steady}).
As the stochastic force further increased to $\sigma_{x,y,z}^2 \sim{\cal O}(1)$, the attractors, ${\bf X}_{{\cal F}_{\pm}}$, become unstable. The trajectory of Lorenz63 in this regime exhibits the similar `butterfly' pattern in Fig. (\ref{Lz63_pt3}a \& b) as we observe for the supercritical case for $R_a=28$ and $f_{x,y,z}=0$ in Fig. (\ref{Lz63_pt2}a \& b). Here the irregular oscillation of the trajectory about ${\bf X}_{\cal F_{+}}$ and ${\bf X}_{\cal F_{-}}$ is due to the combined force of the nonlinear interactions and the stochastic force, $f_{x,y,z}$. Shown in Fig. (\ref{Lz63_pt3})
is the chaotic solution of Lorenz63 for $R_a =20$ and $f_{x,y,z}\sim{\cal N}(0, \ 2)$, where the PDFs, ${\cal P}(x)$, ${\cal P}(y)$ and ${\cal P}(z)$, obtained in DNS are shown in green histograms and the Gaussian distribution with the same mean and variance as for DNS are in dashed red curves. Interestingly, the PDFs of $x$ and $y$ appear to be the superposition of two Gaussian distributions centred at each of the attractor at $\pm 7.1$ and the PDF of $z$ is slightly asymmetric but close to Gaussian. Closely analysing the data, we find that the mean trajectory of $z$ converges to the value, $\langle z \rangle=17.2$, other than the one in the steady state at $R_a-1 = 19$. The Lorenz63 system comprise of two `strange attractors' behaves similarly for different $R_a$ and $f_{x,y,z}$ in the chaotic state.

\subsection{Direct statistical simulation of Lorenz63 in the chaotic states}
\label{Lz63_dss}

We integrate the cumulant equations, starting from the random initial conditions, forwards in time to obtain the statistical equilibrium of the Lorenz63 system in the chaotic states. The results are summarised, and compared with the statistics accumulated from DNS in Table (\ref{Lz63chaotic}).
\begin{table*}
\begin{ruledtabular}
\begin{tabular}{l|c|c|c|c|c|c|c|c|c|c|c|c}
      & $R_a$ & $f_{x,y,z}$  &$\tau_d$ & $C_x$  & $C_y$  & $C_z$    & $C_{xx}$ & $C_{xy}$ & $C_{xz}$ & $C_{yy}$   & $C_{yz}$ & $C_{zz}$ \\\hline
DNS &$28$ & $0$           &              & $0$ & $0$  & $25.10$  & $67.84$ & $66.95$ & $0$ & $84.02$ & $0$ & $58.73$ \\\hline
CE2.5& &$0$           &  $1/40$    & $0$ & $0$  & $23.84$  & $63.58$ & $63.58$ & $0$ & $62.84$ & $0$ & $75.54$ \\\hline
CE2.5& &$0$           &  $1/30$    & $0$ & $0$  & $23.84$  & $66.21$ & $66.21$ & $0$ & $65.93$ & $0$ & $54.01$ \\\hline
CE3 & &$0$           &  $1/20$    & $0$ & $0$  & $23.99$  & $63.96$ & $63.96$ & $0$ & $72.50$ & $0$ & $69.09$  \\\hline
CE3 & &$0$           &  $1/30$    & $0$ & $0$  & $24.54$  & $65.45$ & $65.45$ & $0$ & $71.52$ & $0$ & $58.00$  \\\hline
CE3 & &$0$           &  $1/200$  & $0$ & $0$  & $26.38$  & $70.34$ & $70.34$ & $0$ & $70.80$ & $0$ & $16.23$  \\\hline
CE3 & &$0$           &  $1/15$  & $0$ & $0$  & $23.59$  & $62.90$ & $62.90$ & $0$ & $73.28$ & $0$ & $76.61$  \\\hline
CE3s& &$0$           &  $1/40$    & $0$ & $0$  & $24.92$  & $66.45$ & $66.45$ & $0$ & $70.99$ & $0$ & $50.13$ \\\hline\hline
DNS & $28$ &${\cal N}(5,\ 20)$   &        & $1.78$      & $1.28$     & $17.51$  & $40.95$ & $39.44$ & $8.14$         & $48.10$   & $2.97$         & $37.76$ \\\hline
CE2.5& &${\cal N}(5,\ 20)$   &   $1/44.5$     & $2.06$      & $1.56$     & $24.81$  & $59.97$ & $57.97$ & $10.00$         & $61.48$   & $-2.50$         & $67.05$ \\\hline
CE3 & &${\cal N}(5,\ 20)$           &  $1/10$    & $2.25$ & $1.75$  & $24.75$  & $58.06$ & $57.06$ & $10.57$ & $67.27$ & $-1.19$ & $58.81$ \\\hline\hline
DNS & $20$ &${\cal N}(0, \ 2)$     &    & $0$      & $0$     & $16.68$  & $45.06$ & $44.46$ & $0$         & $53.89$   & $0$         & $44.74$ \\\hline
CE2  & &${\cal N}(0, \ 2)$    &    & $0$      & $0$     & $19.04$  & $50.96$ & $50.76$ & $0$         & $50.93$   & $0$         & $0.75$ \\\hline
CE2.5 & &${\cal N}(0, \ 2)$    & $1/50$   & $0$      & $0$     & $17.25$  & $46.20$ & $46.00$ & $0$         & $46.02$   & $0$         & $31.66$ \\\hline
CE3  & &${\cal N}(0, \ 2)$    & $1/30$   & $0$      & $0$     & $17.14$  & $45.90$ & $45.70$ & $0$         & $48.30$   & $0$         & $32.46$
\end{tabular}
\end{ruledtabular}
\caption{\label{Lz63chaotic}The low-order statistics of Lorenz63 in the chaotic state for $R_a=28$ and $20$ with different stochastic force, $f_{x,y,z}$, where the case labelled as `CE3s' is the solution of the CE3 equation without in the temporal components of third order equations.}
\end{table*}
We observe that the chaotic dynamics of Lorenz63 can be accurately described by the CE3 approximation for a range of eddy damping parameter, $\tau_d$. For Lorenz63 system, the skewness (third order cumulant) that quantifies the asymmetry of the probability distribution is important for accurately approximating the statistical equilibrium of Lorenz63. For example, the second order cumulant, $C_{zz}$ is purely determined by the third order term, $C_{xyz}$, see Eq. (\ref{Lz63_ce2}). The most accurate solutions of CE3 equations are obtained for $\tau_d$ in the range between ${{\cal O}(10^{-2})}$ and ${{\cal O}(10^{-1})}$, which is approximately $10$ to $100$ times smaller than $t_c$ or $t_w$.

The CE2.5 approximation is also found numerically stable for all test cases for the eddy damping parameter, $\tau_d$ in the same range as for CE3 equations from ${\cal O}(10^{-2})$ to ${\cal O}(10^{-1})$. This approximation assumes that the terms involving the first order cumulants, $C_{x}$, $C_{y}$ and $C_{z}$, are statistically insignificant in the governing equation of the third order and are neglected in the numerical computation. The solution of the CE2.5 approximation is found to be as accurate as CE3.

For all cases listed in Table (\ref{Lz63chaotic}), the CE2 approximation does not converge, except for the case for $R_a=20$ and $f_{x,y,z}\sim {\cal N}(0, \ 2)$. However, for this parameter set, the CE2 approximation very poorly estimates the statistical equilibrium, e.g., the first cumulant, $C_{z}$, fails to converge to the mean trajectory of $z$ in the chaotic state but to the one in the steady state at $R_a-1=19$; the second cumulant, $C_{zz}$, is inaccurately determined due to the lack of the knowledge of the third cumulant, $C_{xyz}$.

\subsection{Fixed points for Lorenz63 in the chaotic state}

Although timestepping allows the access of the stable solutions of the cumulant equations, other fixed points are possible solutions as discussed earlier. Here we assess the effectiveness of various methods for accessing these fixed points.

\begin{figure*}
\centering
\subfigure[The fixed point of the attractor as a function of $R_a$]
{
	\includegraphics[width=0.3\hsize]{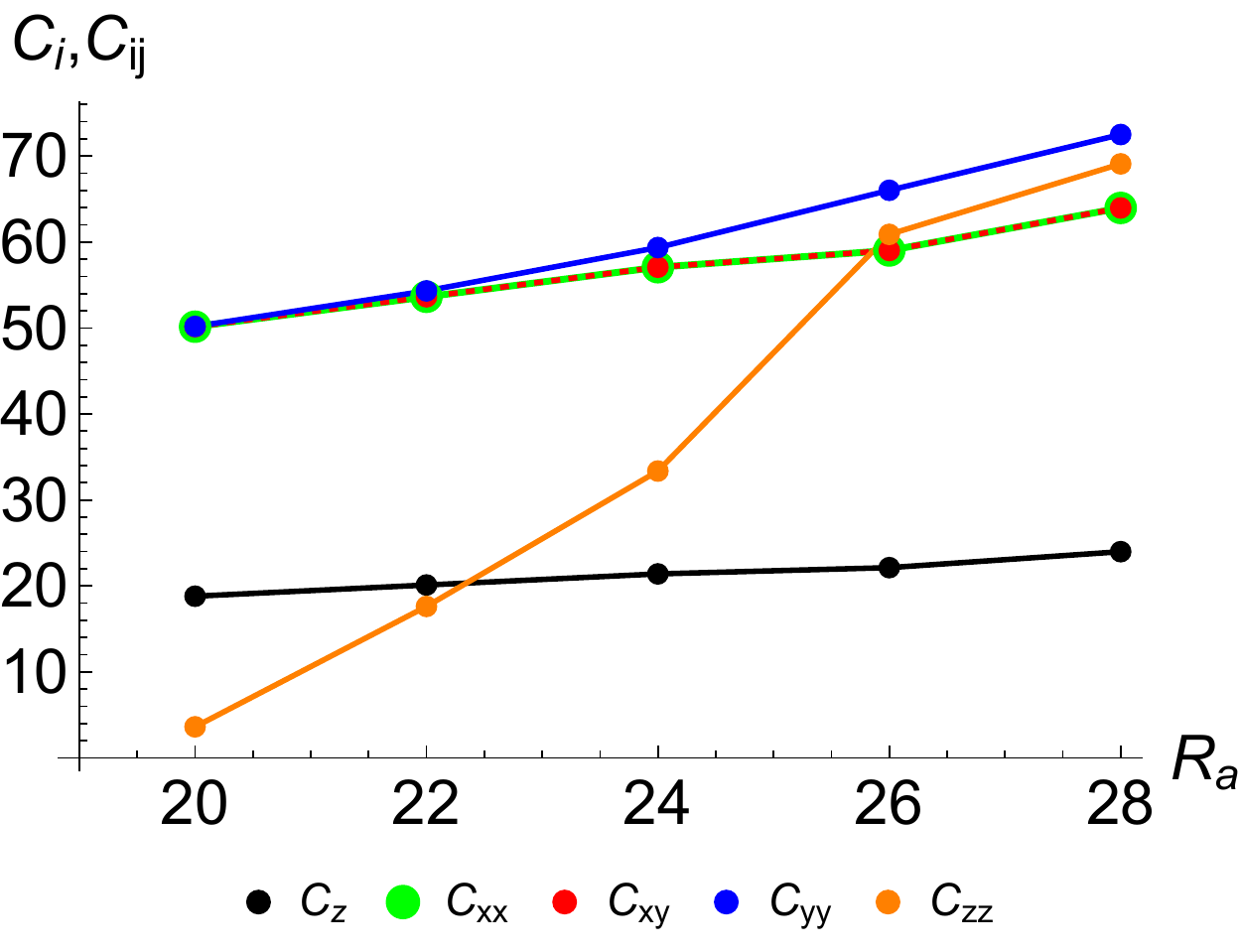}
}
\subfigure[The fixed point of the attractor as a function of $\tau_d$]
{
	\includegraphics[width=0.3\hsize]{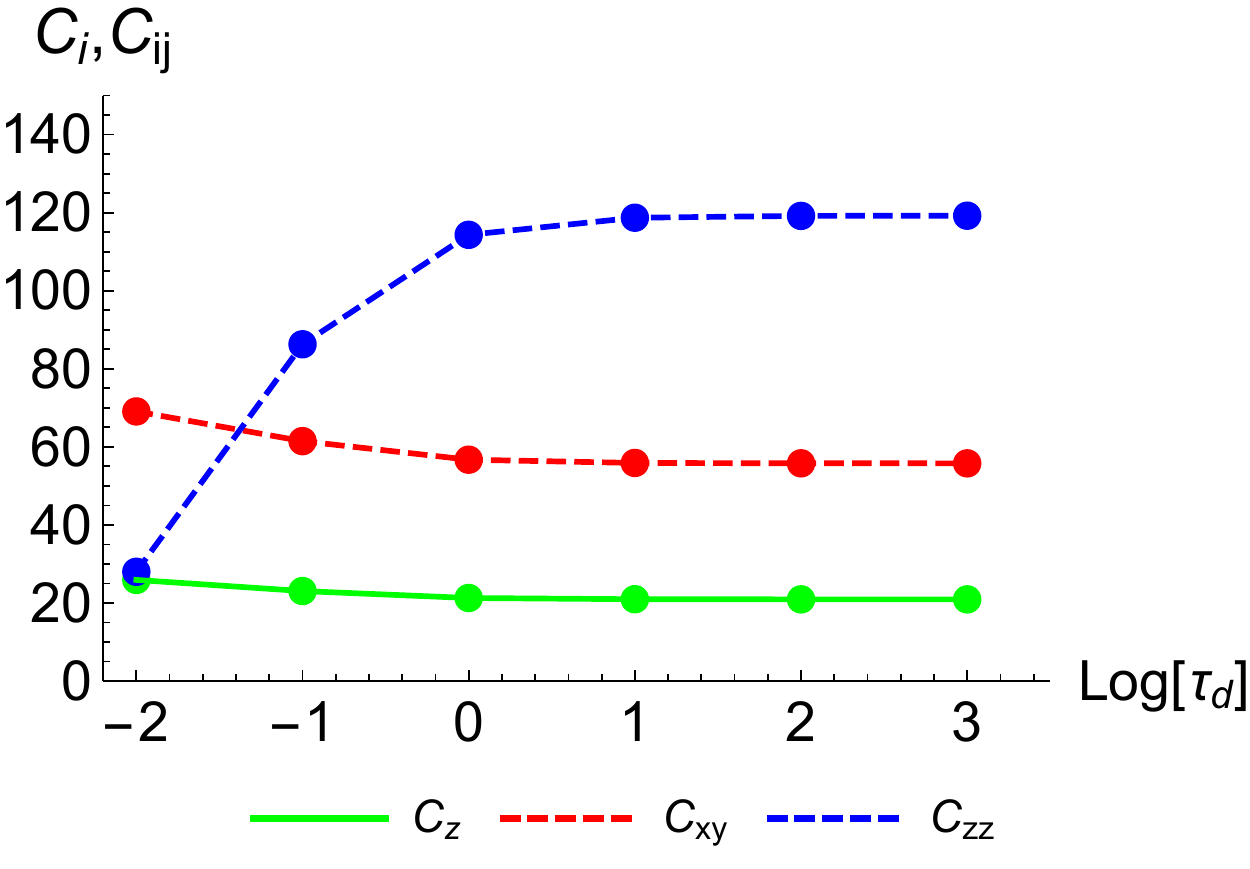}
}
\subfigure[An unstable fixed point as a function of $\tau_d$]
{
	\includegraphics[width=0.3\hsize]{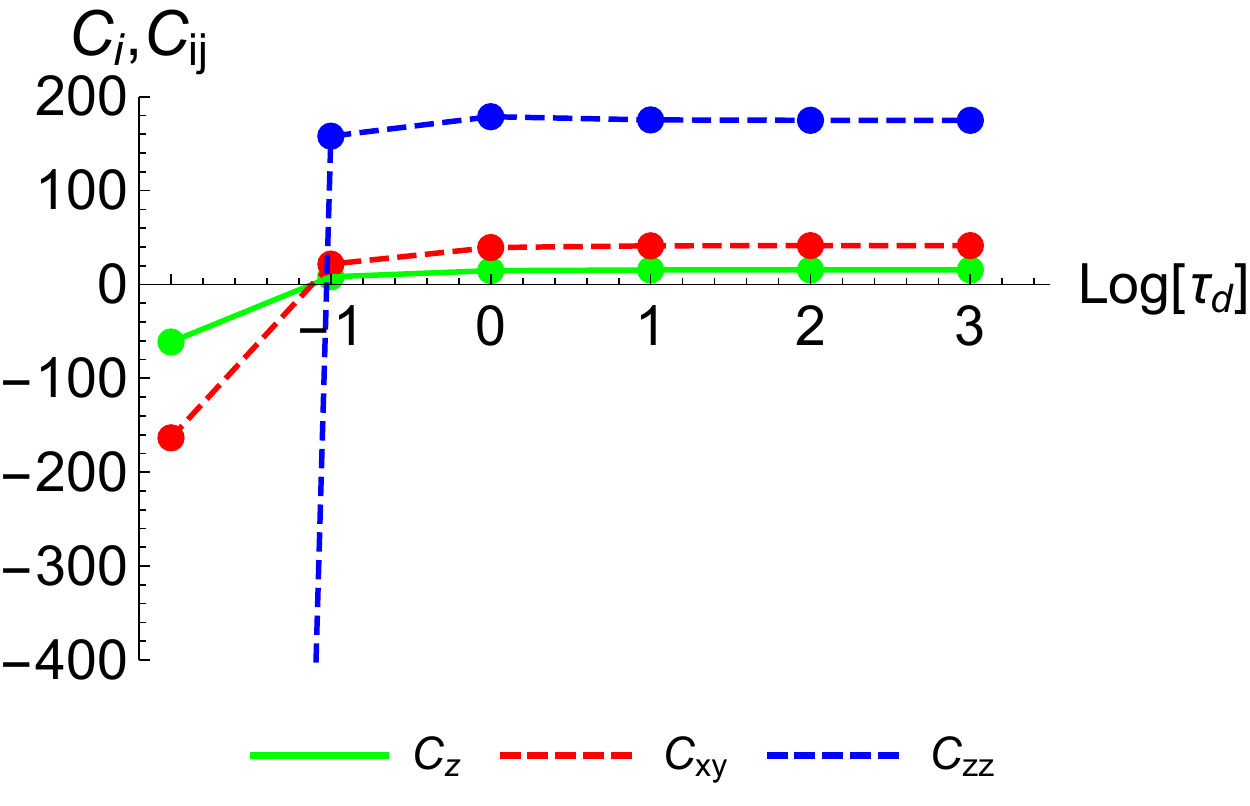}
}
\caption{The illustration of the continuation of the attractor as a function of Rayleigh number, $R_a$, from $R_a=20$ to $30$ in (a) and as a function of the eddy damping parameter, $\tau_d$ in (b); the continuation of an unstable attractor as a function of $\tau_d$ in (c), where the unstable fixed point becomes statistically realisable as $\tau_d>0.1$.}
\label{CumRa}
\end{figure*}

The fixed points of Lorenz63 can be solved directly via the symbolic packages of {\it Mathematica} or {\it Python}, where the fixed points are assumed invariant in time. We obtain $7$ statistically realizable fixed points out of $41$ roots of CE3 equations of Lorenz63 defined in (\ref{Lz63_ce1}, \ref{Lz63_ce2} \& \ref{Lz63_ce3}) for $R_a=28$, $f_{x,y,z}=0$ and $\tau_d^{-1} = 20$. The fixed point that corresponds to the strange attractors is the only stable one in time. As increasing the dynamics of Lorenz63 from the steady state to the chaotic state by increasing the Rayleigh number, the stable attractors evolves to the `strange' attractors, see Fig. (\ref{CumRa}a)
for the continuation of the attractor as a function of Rayleigh number for $f_{x,y,z}=0$. If $\tau_d$ is small, the third order equations are statistically insignificant in the CE3 equations and the CE3 approximation is reduced to CE2, see Fig. (\ref{CumRa}b) for the continuation of the attractor as a function of the eddy damping parameter, $\tau_d$. We also observe that other fixed points are sensitive to the choice of $\tau_d$. Shown in Fig. (\ref{CumRa}c) is the solution of an unstable attractor of Lorenz63 as a function of $\tau_d$, where the second order cumulant, $C_{zz}>0$, is statistically realisable for $\tau_d > 0.1$ and becomes non-realisable ($C_{zz}<0$) when $\tau_d < 0.1$.

\begin{figure*}[htp]
\centering
\subfigure[]
	{
		\includegraphics[width=0.25\hsize]{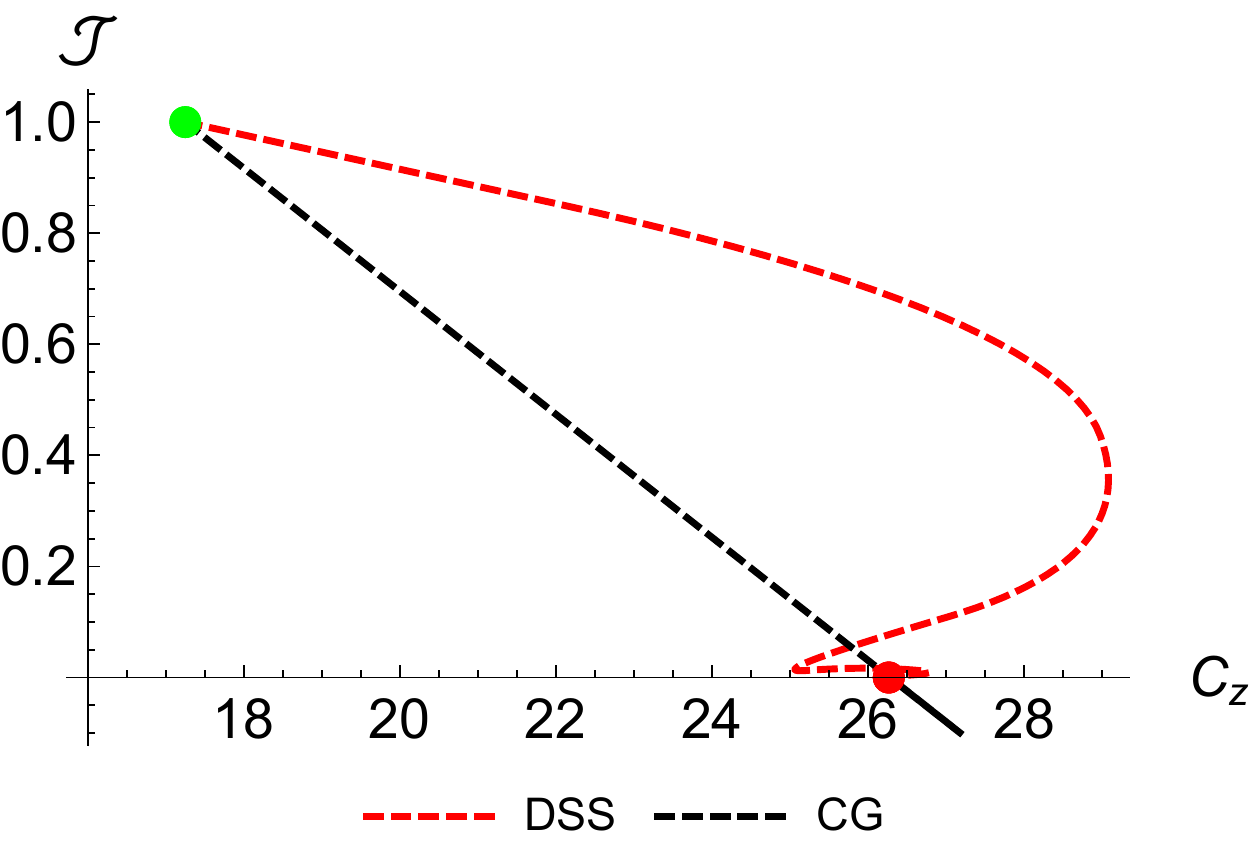}
	}
\subfigure[]
	{
		\includegraphics[width=0.25\hsize]{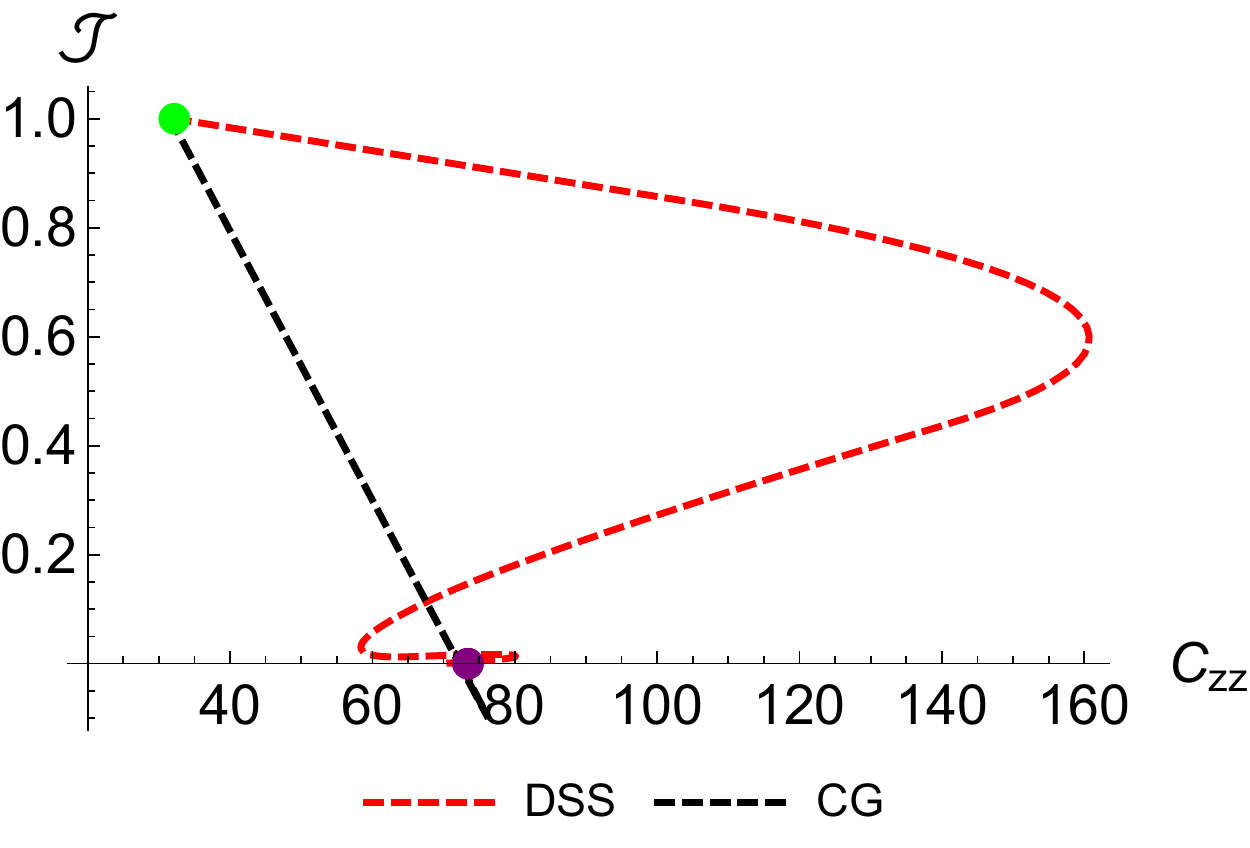}
	}
\subfigure[] 
	{
		\includegraphics[width=0.25\hsize]{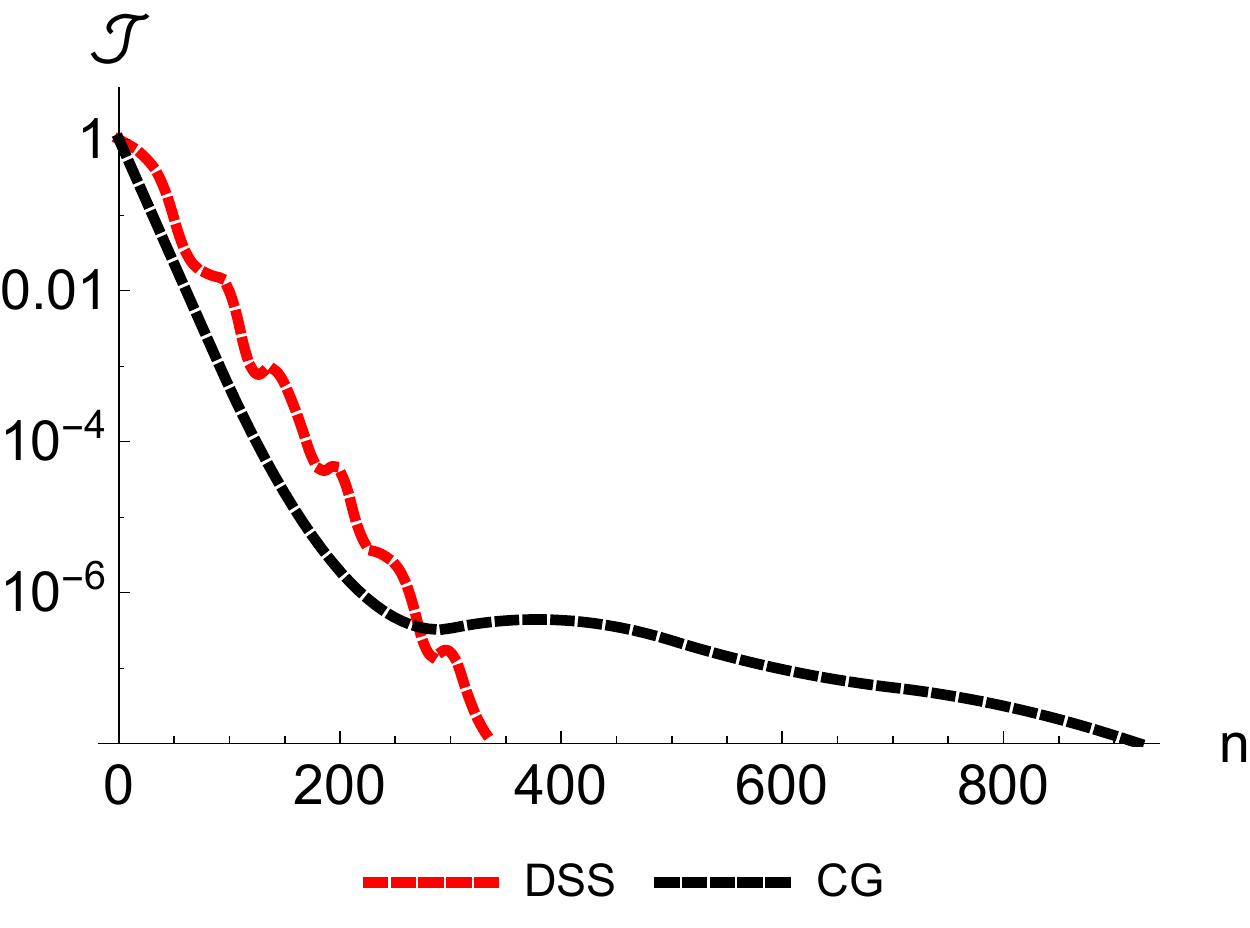}
	}
\caption{The stable fixed point is found by minimizing ${\cal J}$ via the {\it CG} and time stepping method for the CE2.5 approximation of the Lorenz63 system for $R_a=30$ and $f_{x,y,z} \sim{\cal N}(0, \ 2)$, where (a) and (b) show the path of the cumulants, $C_z$ and $C_{zz}$, as the solution converges to ${\cal J} = 0$ and (c) illustrates the convergence of ${\cal J}$ as a function of the number of iterations, $n$. The initial guess and the terminal solution of the minimization are shown as green and red dots in (a) and (b). The optimal solution converges to the stable time-invariant solution of CE2.5 via the time stepping method.}
\label{Lz63_pt6}
\end{figure*}

We also study gradient based optimization methods for computing the fixed point of the CE2.5 approximations, in the presence of noise. For $\sigma_{x,y,z}^2 = 2$, the dynamical system can be accurately approximated by timestepping the CE2.5 equations and the optimal solution for ${\cal J}$ converges to the same fixed point as obtained via timestepping. Shown in Fig. (\ref{Lz63_pt6})
is the convergence to the optimal solution for the cumulants of the CE2.5 approximation of the Lorenz63 in the chaotic state for $R_a = 38$ and $f_{x,y,z} \sim {\cal N}(0, 2)$ via the {\it CG} method, where ${\cal J}$ is normalized by its value at the first iteration. The initial guess is taken from the fixed point for $R_a=28$ and $f_{x,y,z} \sim {\cal N}(0, 2)$. It is of importance to note that the misfit, ${\cal J}$, of Lorenz63 is not convex everywhere. If the initial guess is randomly chosen, the optimal solution of the CE2.5 system may converge to other fixed points. But we observe that for CE2.5 approximation the stable fixed point can always be found by continuing the stable fixed point from the solutions of the nearby control parameter. For the first ${\cal O}(300)$ iterations, the {\it CG} method slightly outperform the time stepping method in terms of the convergence rate, where the misfit, ${\cal J}$, reduces by a factor of $10^{7}$ and the optimal time step, $dt \simeq 10^{-2}$, is used for the numerical integration. After approximately $300$ iterations, the convergence rate of ${\cal J}$ of {\it CG} flattens out but the rate for timestepping remains constant. For this case the quasi-newton method performs similarly as {\it CG} in terms of the accuracy and the convergence rate. We note that calculating the `downhill' direction for minimizing the misfit, ${\cal J}$, that is directly computed via the symbolic differentiation is almost computationally as expensive as for evolving the cumulant equation one time step forward.

We find that it is difficult to find the stable fixed point for CE3 approximations via the gradient based method. The optimization either converges slowly to unstable or non-realizable fixed points or is trapped by the local minima that are introduced by the third order cumulants.

\begin{figure*}[htp]
\centering
\subfigure[]
	{
		\includegraphics[width=0.25\hsize]{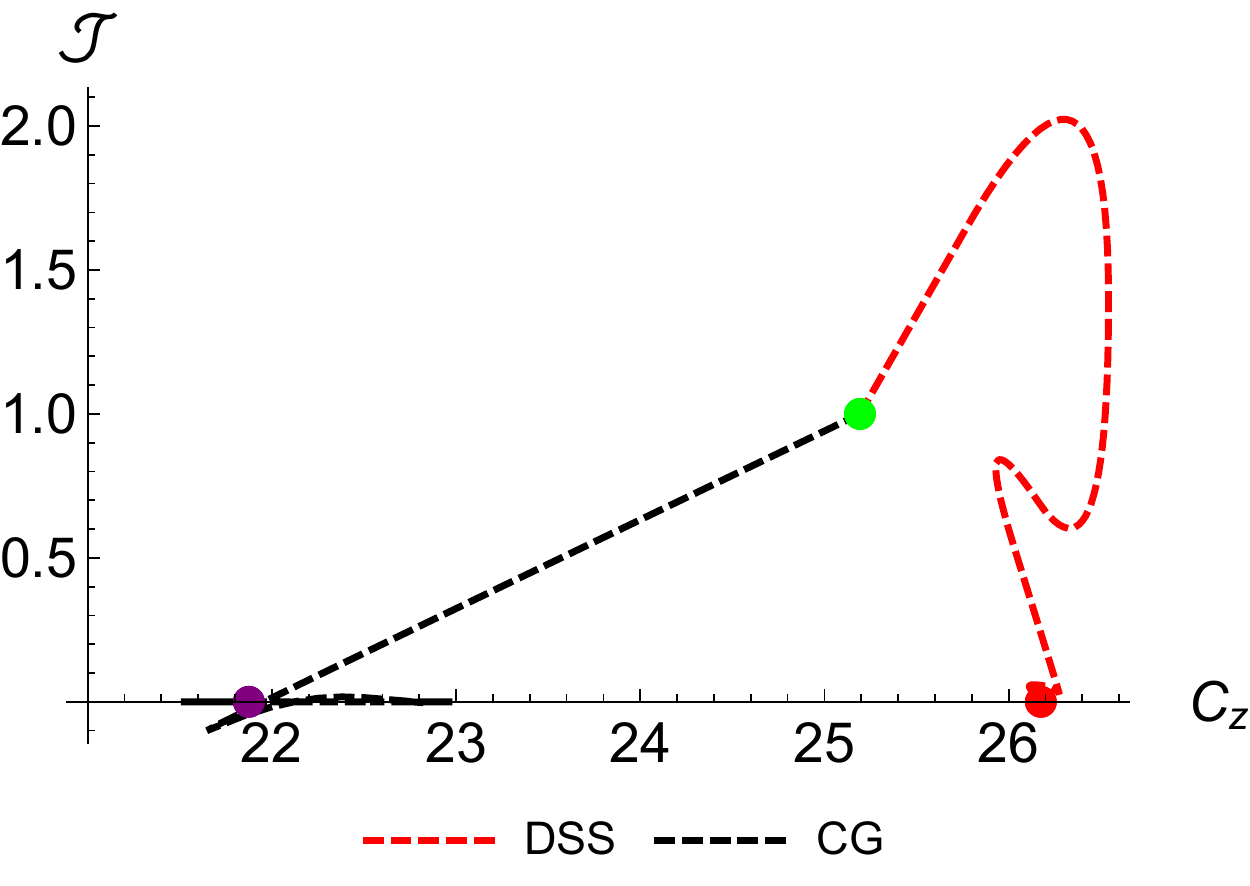}
	}
\subfigure[]
	{
		\includegraphics[width=0.25\hsize]{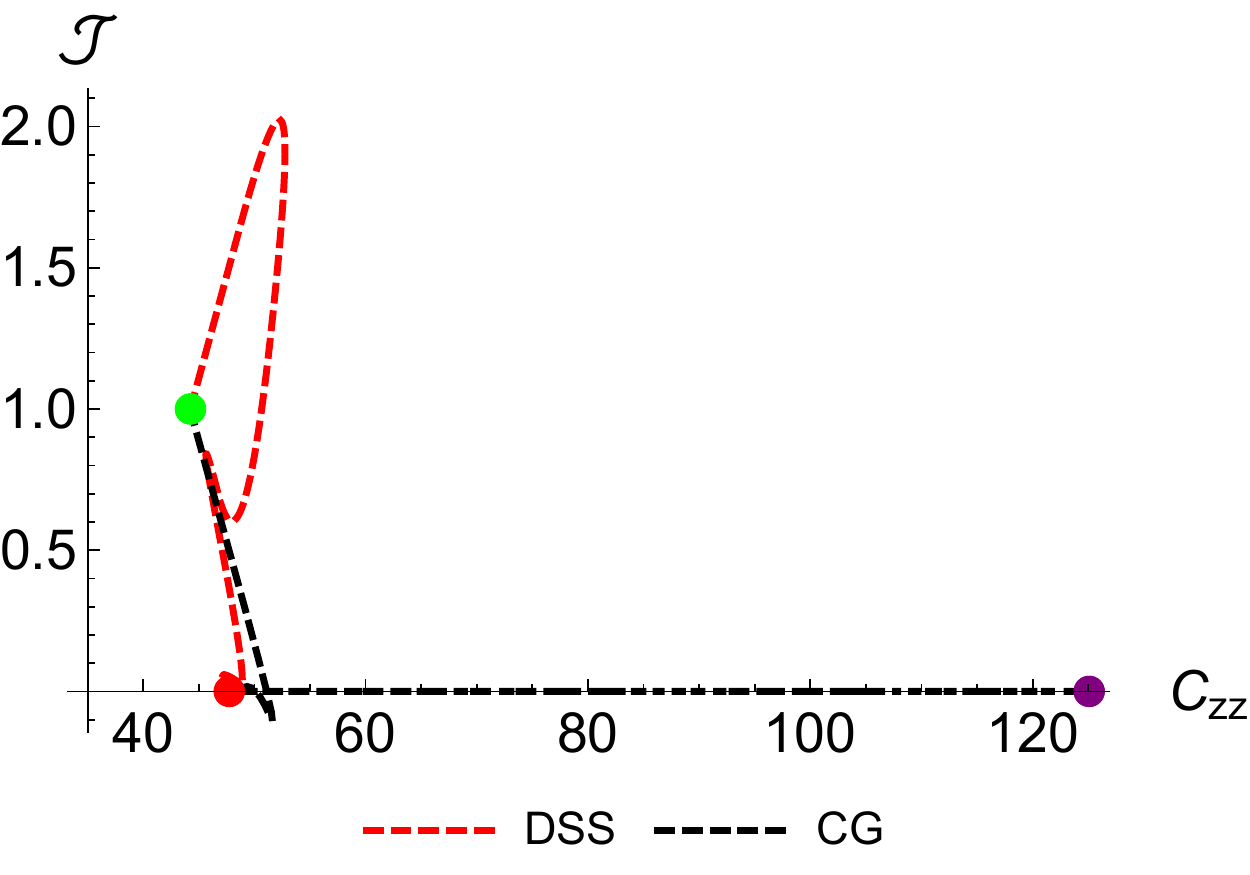}
	}
\subfigure[] 
	{
		\includegraphics[width=0.25\hsize]{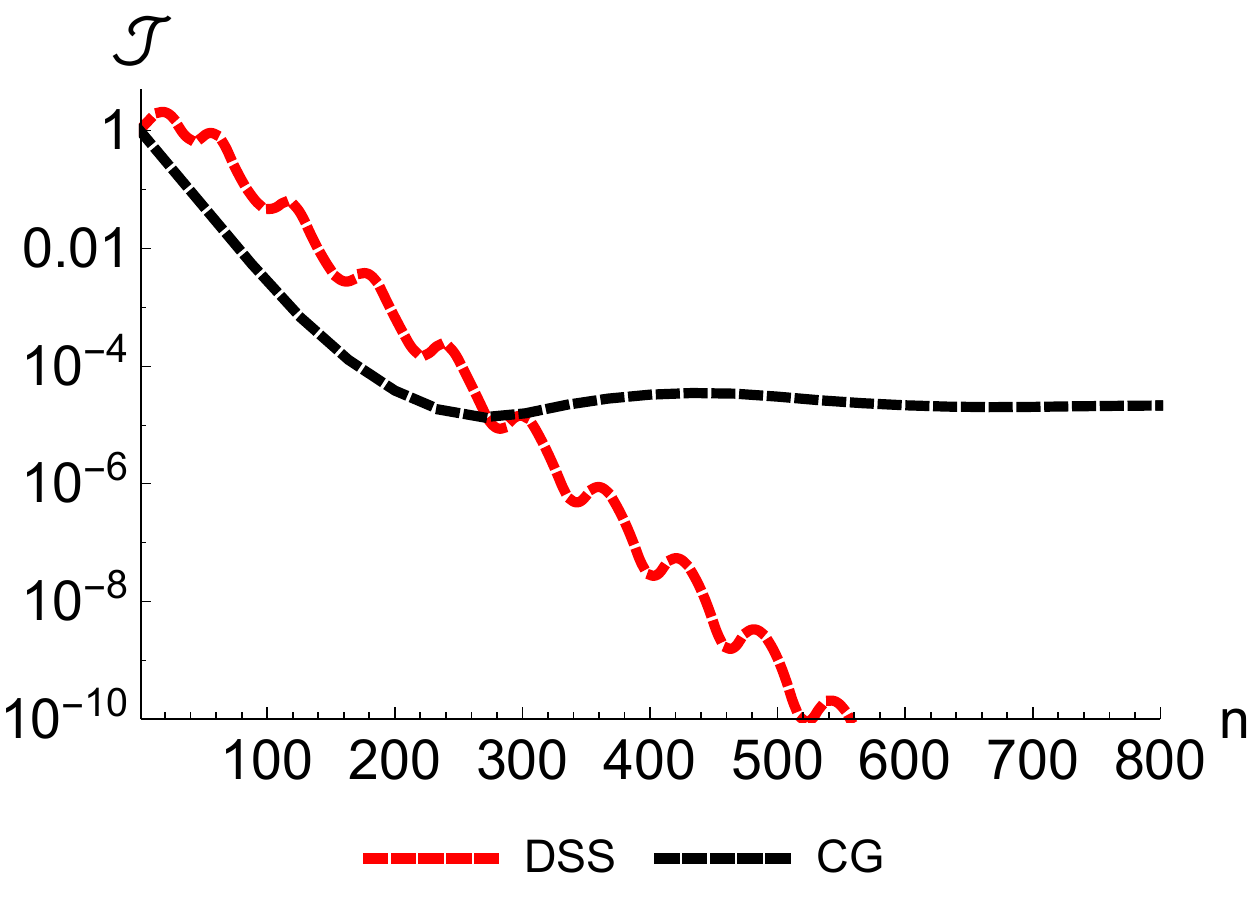}
	}
\caption{The migration of $C_z$ and $C_{zz}$ from $R_a = 28$ to $29$ via the time stepping and {\it CG} method for CE3 equations in (a) \& (b) and the convergence of ${\cal J}$ as the number of time steps/iterations in (c), where the stochastic force is zero, i.e., $f_{x,y,z} = 0$, the green dots in (a) and (b) stand for the initial condition/guess for the time stepping/CG method, the red dots are for the time invariant solution found by time stepping method and the purple ones are for the terminal solution of {\it CG}.}
\label{Lz63_pt7}
\end{figure*}

As an example, we use {\it CG} to continue the stable fixed point of Lorenz63 from $R_a = 28$ (found by timestepping) to $R_a=29$ with $\Delta R_a = 1$, for the same stochastic force level ($f_{x,y,z} \sim {\cal N}(0, 2)$). The results are in Fig. (\ref{Lz63_pt7})
which shows the path of the low-order cumulant, $C_z$ and $C_{zz}$, as ${\cal J}$ converges to zero and the convergence of ${\cal J}$ as the number of iterations, where the green and purple dots in Figs. (\ref{Lz63_pt7}a \& b) represent the initial guess and terminal solution of the optimization and the red dots are for the solution found by timestepping. Disappointingly, the optimization goes into the wrong direction and is trapped by the local minima, where the misfit saturate at the level of ${\cal J}\sim{O}(10^{-4})$. Similar performance is observed for different $R_a$ and $\Delta R_a$ with and without the stochastic force, $f_{x,y,z}$.

\begin{figure*}[htp]
\centering
\subfigure[$f_{x,y,z}=0$ for CE3]
	{
		\includegraphics[width=0.25\hsize]{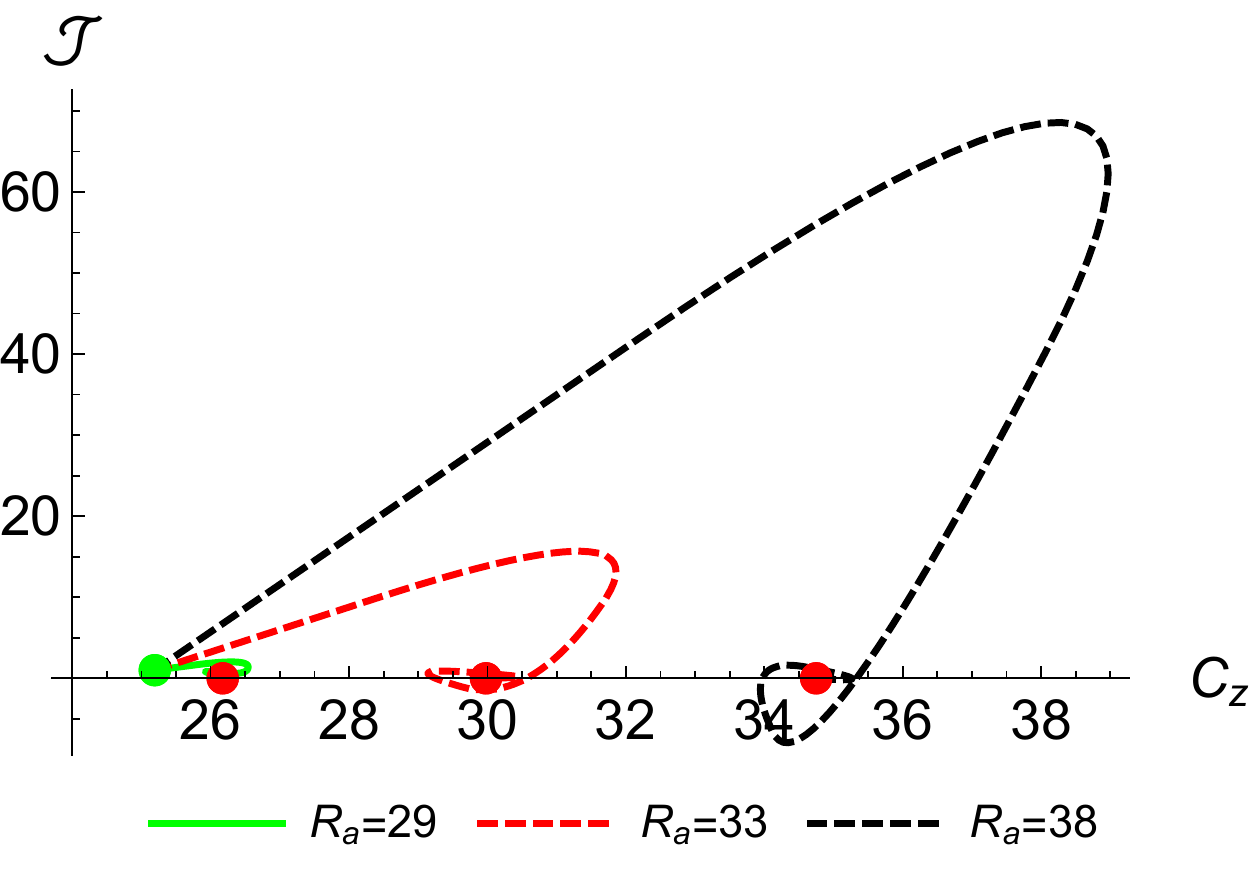}
	}
\subfigure[$f_{x,y,z}=0$]
	{
		\includegraphics[width=0.25\hsize]{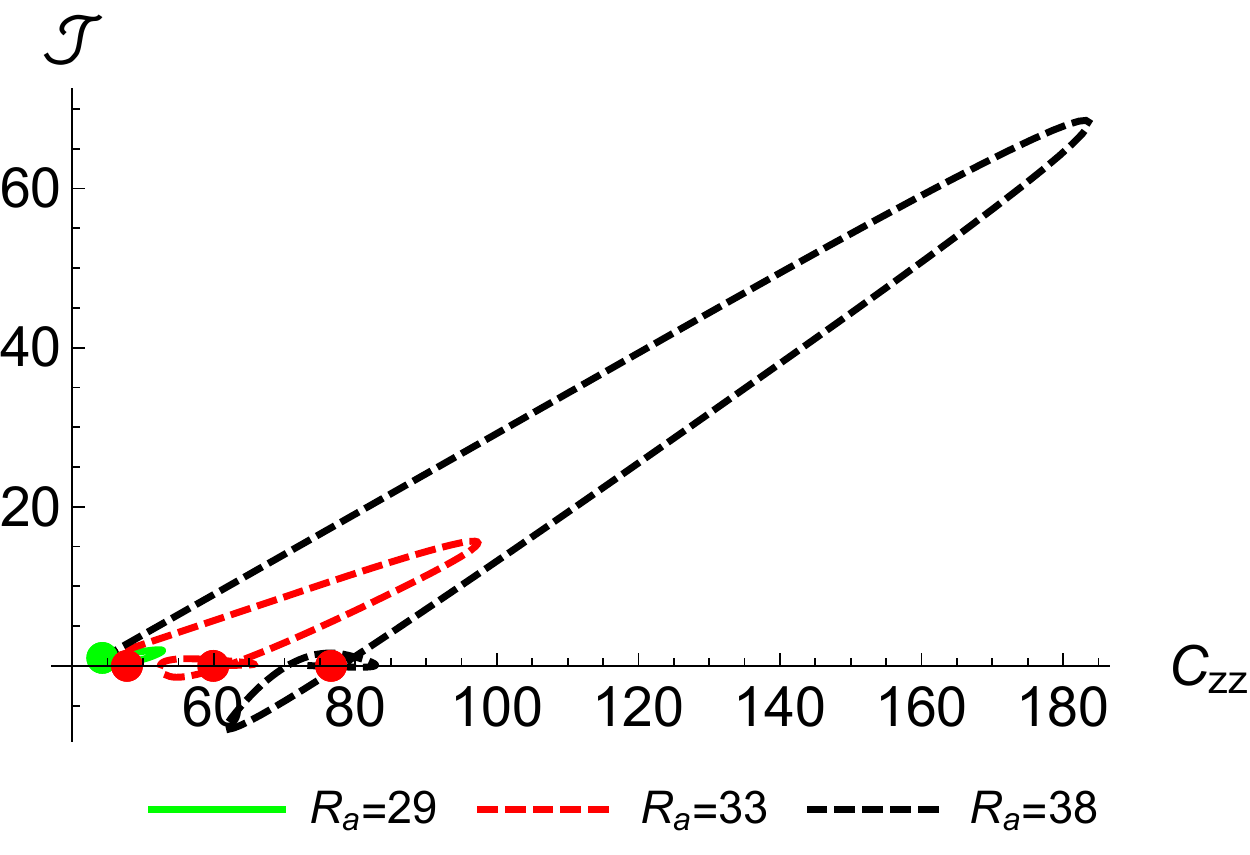}
	}
\subfigure[$f_{x,y,z}=0$] 
	{
		\includegraphics[width=0.25\hsize]{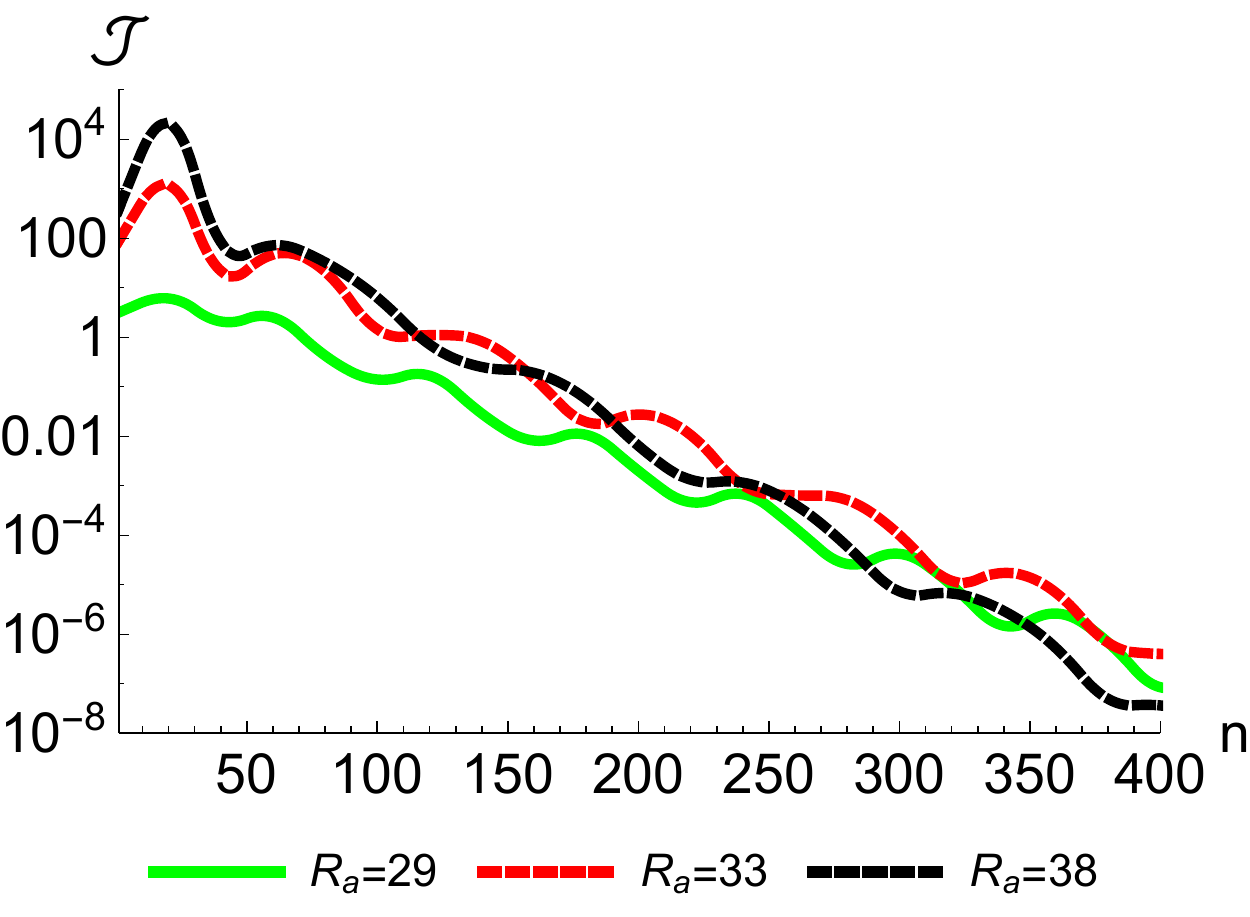}
	} \\
\subfigure[$f_{x,y,z}\sim{\cal N}(0,\ 10)$]
	{
		\includegraphics[width=0.25\hsize]{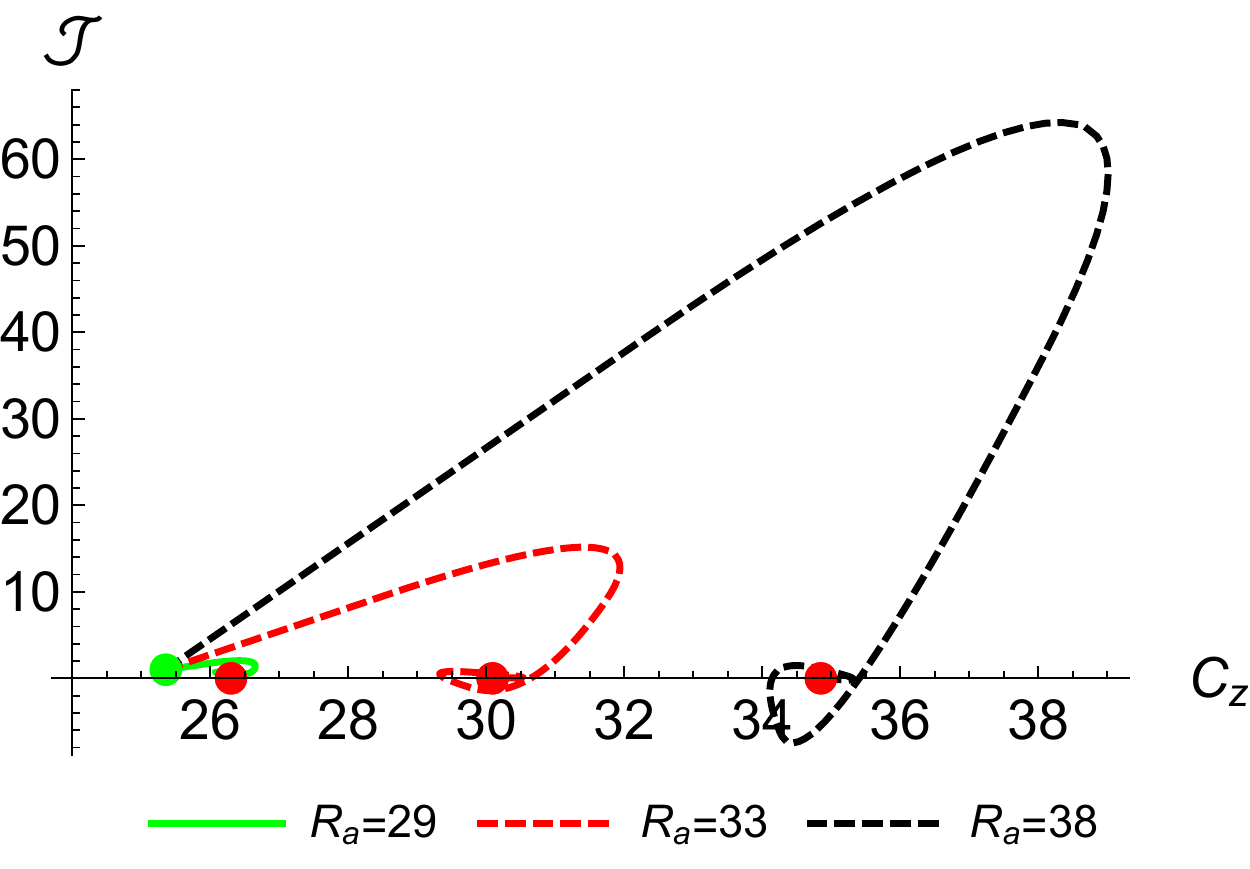}
	}
\subfigure[$f_{x,y,z}\sim{\cal N}(0,\ 10)$]
	{
		\includegraphics[width=0.25\hsize]{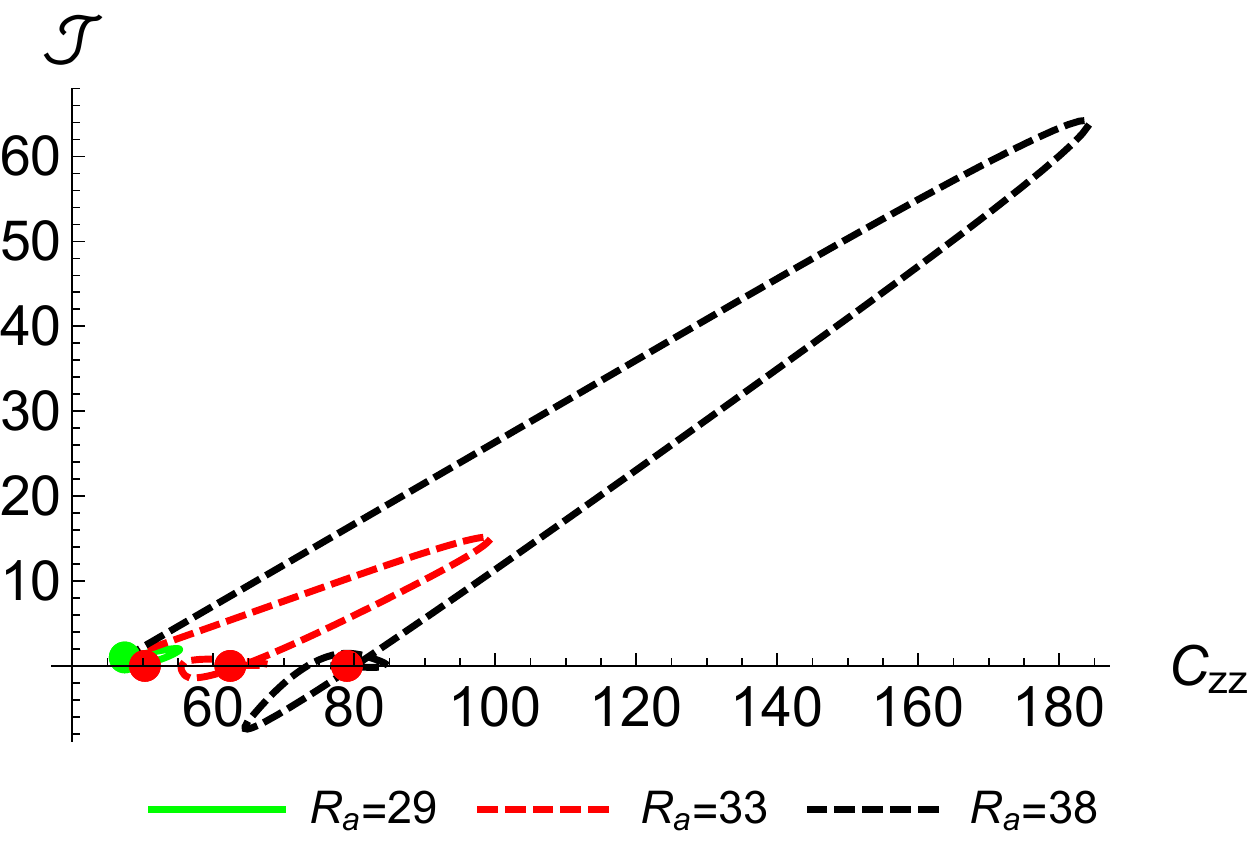}
	}
\subfigure[$f_{x,y,z}\sim{\cal N}(0,\ 10)$] 
	{
		\includegraphics[width=0.25\hsize]{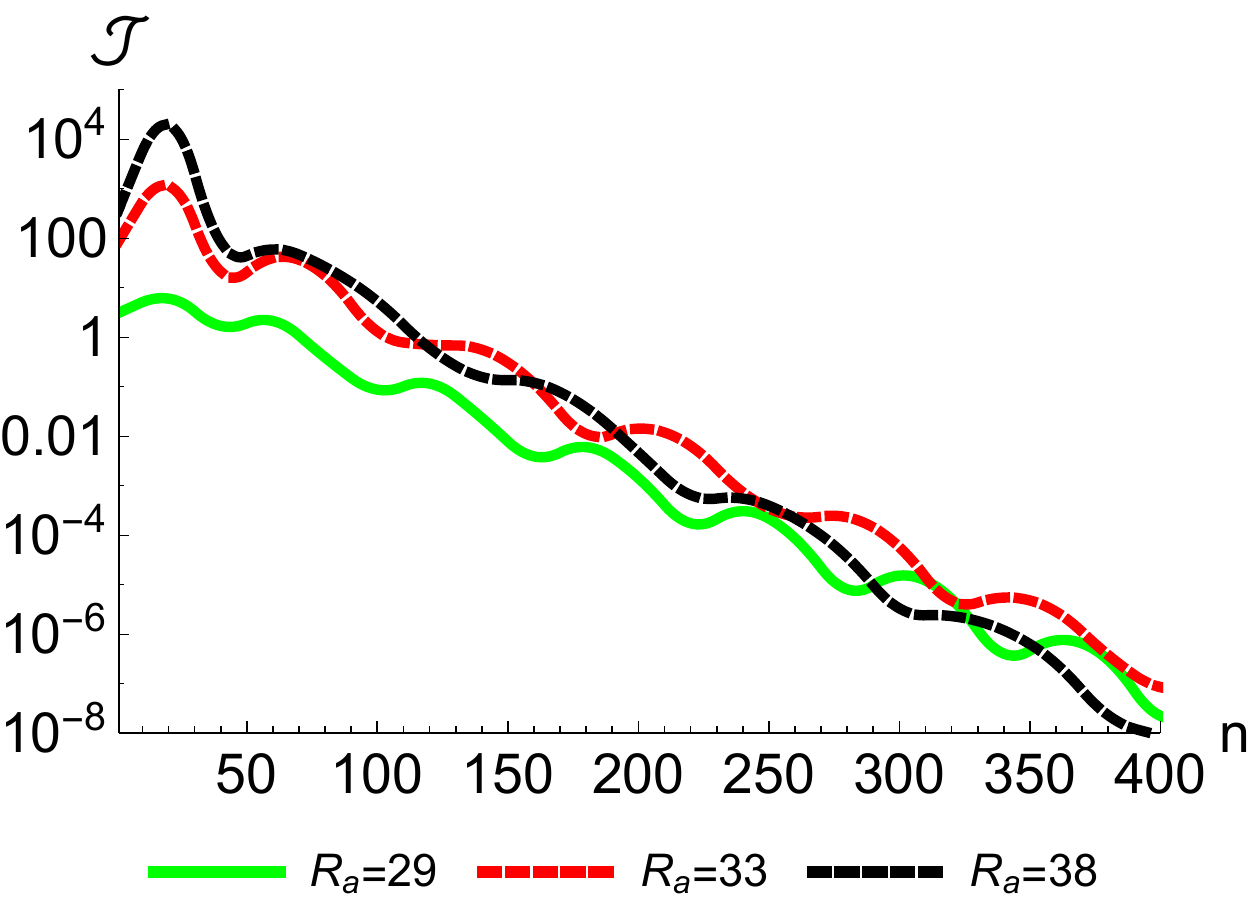}
	}
\caption{The illustration of the fixed points of the CE3 approximation of the Lorenz63 system in the chaotic state and for different $R_a$ found by the time stepping method, where (a)--(b) and (d)--(e) show the path of $C_z$ and $C_{zz}$ as ${\cal J}$ tends to zero and (c) and (f) show the exponential convergence of ${\cal J}$ as the number of iterations.}
\label{Lz63_pt4}
\end{figure*}

The stable, time-invariant solution of the CE2.5/3 approximations of Lorenz63 in the chaotic state can always be obtained by the time stepping method. An interesting calculation is to determine the path of approach to a fixed point using a nearby solution as an initial guess. Shown in Fig. (\ref{Lz63_pt4})
are the paths of the low-order cumulants, $C_z$ and $C_{zz}$, as ${\cal J}$ approaches to zero and the convergence of ${\cal J}$ as the number of time steps for the CE2.5/3 approximation of Lorenz63. Here the initial state is that calculated at $R_a = 28$ for $f_{x,y,z}=0$ and $f_{x,y,z}\sim{\cal N}(0, \ 10)$ and we show 3 different cases with $\Delta R_a = 1, 5$ and $10$. Interestingly, the path of the cumulant, e.g., $C_z$ and $C_{zz}$, demonstrates a strong self-similarity as we evolve the CE3 equations in time to reduce ${\cal J}$ for different $\Delta R_a$. The form of the approach to the fixed point suggests that taking reasonably large values of $\Delta R_a$ is the best strategy for continuation of solutions. For the cumulant system with non-negligible third order cumulant for $f_{x,y,z} = 0$, the misfit, ${\cal J}$, is never found convex as the solution of the cumulant equations converges to the stable fixed point. The same observations have been made for all other test cases. We speculate that this is the reason why the minimization method fails to optimize the CE3 system.

\section{Conclusion}
\label{conclusion}

In this paper, we implement direct statistical simulation to study the simplified highly nonlinear dynamical system, Lorenz63, and apply DSS to study the statistical behavior of this system in the chaotic regime without and with external random forcing.

We find that the CE2.5 and CE3 approximations are sufficiently accurate to describe the long-term statistical evolution of these systems, though the probability distributions of these systems are either strongly asymmetric or have longer tails than a Gaussian distribution. The mean trajectory is very accurately determined by the cumulant equations with a maximum truncation at third order in the cumulants. There is a relatively small error of less than 1\% in the mean statistics compared with those obtained via the traditional approach of long time averaging of direct numerical simulation. The interactions between the coherent and non-coherent components of the dynamics are also accurately quantified by the second order terms with relative errors of less than 20\%. For both of the CE2.5 and CE3 closures a single eddy damping parameter, $\tau_d$, is introduced to approximate the correlation time among the non-coherent components of the dynamics and to stabilize the numerical integration of DSS equations. The optimal $\tau_d$ is found to be approximately $10$ to $100$ times smaller than the characteristic time scale of the Lorenz63 system, $t_w$ and $t_c$, in the range of $10^{-2}$ to $10^{-1}$. This result is consistent with \cite{allawala_2016}.

We also attempt to directly access the fixed points of the cumulant equations of Lorenz63 systems. For this dynamical system, all time invariant solutions can be solved symbolically. There are 7 roots among the 41 solutions of Lorenz63 that are statistically realizable for $\tau_d = 1/20$ but the fixed point corresponding to the strange attractor is the only one stable in time. The same symbolic technique cannot be applied to discover the fixed points of systems with large numbers of degree of freedoms, due to the rapid increase in computational complexity. We also find that the time stepping method and the gradient methods converge to the statistical equilibrium exponentially at comparable rates, except in the case of CE3 for which the gradient approach never converges. 

In conclusion we have demonstrated the effectiveness of the cumulant equations to describe the statistical evolution of the highly nonlinear dynamical system that represent dimensionally reduced fluids.  We note that turbulent fluid dynamical systems with a huge number of freedoms are of course governed by partial differential equations are typically less chaotic but more diffusive in the statistical space. It may be interesting to investigate the use of gradient based methods in conjunction with DSS to find the steady-state statistics of such fluids.

\section*{Acknowledgements}

This is supported in part by European Research Council (ERC) under the European Unions Horizon 2020 research and innovation program (grant agreement no. D5S-DLV-786780) and by a grant from the Simons Foundation (Grant number 662962, GF). 

\bibliographystyle{plainnat}
\bibliography{Lorenz}

\appendix
\section{The third order cumulant equations of Lorenz63}
\label{Lz63_CE3}

The third order cumulant equations for the CE3 approximation of the Lorenz63 system consists of ten equations, i.e.,
{\small
\begin{widetext}
\begin{eqnarray}
\left(d_t +3 P_r +\frac{1}{\tau_d}\right)C_{xxx} &=&   3P_r C_{xxy}  \nonumber\\
\left(d_t +2 P_r+1 +\frac{1}{\tau_d}\right) C_{xxy} &=&   -C_xC_{xxz} - 2C_{xx}C_{xz} - C_{xxx}C_z + R_a C_{xxx} + 2 P_r  C_{xyy}  \nonumber\\ 
\left(d_t +P_r +2 +\frac{1}{\tau_d}\right) C_{xyy} &=&   -2 C_x C_{xyz} - 2 C_{xx} C_{yz} - 2 C_{xxy} C_z + 2R_aC_{xxy} - 2C_{xy}C_{xz} + P_r C_{yyy}   \nonumber\\
\left(d_t +3+\frac{1}{\tau_d}\right)C_{yyy} &=&   -3 C_x C_{yyz} - 6 C_{xy} C_{yz} - 3 C_{xyy} C_z + 3 R_a C_{xyy}   \nonumber\\ 
\left(d_t +2 P_r + \beta +\frac{1}{\tau_d}\right) C_{xxz} &=&   C_xC_{xxy} + 2C_{xx}C_{xy} + C_{xxx}C_y + 2 P_r C_{xyz}   \nonumber\\
\left(d_t +P_r +\beta + 1+\frac{1}{\tau_d} \right) C_{xyz} &=&   C_x C_{xyy} - C_x C_{xzz} + C_{xx} C_{yy} - C_{xx} C_{zz} + C_{xxy} C_y - C_{xxz} C_z + R_a C_{xxz}   \nonumber \nonumber\\ &&  + C_{xy}^2- C_{xz}^2 + P_r C_{yyz}  \nonumber\\
\left(d_t +\beta + 2 +\frac{1}{\tau_d}\right)C_{yyz} &=&   C_x C_{yyy} - 2C_x C_{yzz} + 2C_{xy}C_{yy} - 2C_{xy}C_{zz} + C_{xyy}C_y - 2C_{xyz}C_z\nonumber\nonumber\\ && + 2 R_a C_{xyz} - 2C_{xz}C_{yz}   \nonumber\\
\left(d_t +P_r +2\beta +\frac{1}{\tau_d}\right) C_{xzz} &=&   2 C_x C_{xyz} + 2 C_{xx} C_{yz} + 2 C_{xxz} C_y + 2 C_{xy} C_{xz} + P_r C_{yzz}  \nonumber\\ 
\left(d_t +2\beta + 1 +\frac{1}{\tau_d}\right)C_{yzz} &=&   2 C_x C_{yyz} - C_x C_{zzz} + 2 C_{xy} C_{yz} + 2 C_{xyz} C_y + 2 C_{xz} C_{yy} - 2C_{xz}C_{zz}  \nonumber \nonumber\\ &&- C_{xzz}C_z + R_a C_{xzz}   \nonumber\\ 
\left(d_t +3\beta +\frac{1}{\tau_d}\right) C_{zzz} &=&   3 C_x C_{yzz} + 6 C_{xz} C_{yz} + 3 C_{xzz} C_y. 
\label{Lz63_ce3}
\end{eqnarray}
\end{widetext}
}

\end{document}